\documentclass[twoside,twocolumn,9pt]{article}
\usepackage{extsizes}
\usepackage[super,sort&compress,comma]{natbib} 
\usepackage[version=3]{mhchem}
\usepackage[left=1.5cm, right=1.5cm, top=1.785cm, bottom=2.0cm]{geometry}
\usepackage{balance}
\usepackage{mathptmx}
\usepackage{sectsty}
\usepackage{graphicx} 
\usepackage{lastpage}
\usepackage[format=plain,justification=justified,singlelinecheck=false,font={stretch=1.125,small,sf},labelfont=bf,labelsep=space]{caption}
\usepackage{float}
\usepackage{fancyhdr}
\usepackage{fnpos}
\usepackage[english]{babel}
\addto{\captionsenglish}{%
  
}
\usepackage{array}
\usepackage{droidsans}
\usepackage{charter}
\usepackage[T1]{fontenc}
\usepackage[usenames,dvipsnames]{xcolor}
\usepackage{setspace}
\usepackage[compact]{titlesec}
\usepackage{hyperref}
\usepackage{amsfonts}
\usepackage{amsmath}
\usepackage{amssymb}
\usepackage{booktabs}
\usepackage{bm}
\usepackage{placeins}
\usepackage[]{algorithm}
\usepackage{algpseudocode}
\usepackage{epstopdf}%This line makes .eps figures into .pdf - please comment out if not required.

\definecolor{cream}{RGB}{222,217,201}

\begin{document}

\pagestyle{fancy}
\thispagestyle{plain}
\fancypagestyle{plain}{
%%%HEADER%%%
\renewcommand{\headrulewidth}{0pt}
}
%%%END OF HEADER%%%

%%%PAGE SETUP - Please do not change any commands within this section%%%
\makeFNbottom
\makeatletter
\renewcommand\LARGE{\@setfontsize\LARGE{15pt}{17}}
\renewcommand\Large{\@setfontsize\Large{12pt}{14}}
\renewcommand\large{\@setfontsize\large{10pt}{12}}
\renewcommand\footnotesize{\@setfontsize\footnotesize{7pt}{10}}
\makeatother

\renewcommand{\thefootnote}{\fnsymbol{footnote}}
\renewcommand\footnoterule{\vspace*{1pt}% 
\color{cream}\hrule width 3.5in height 0.4pt \color{black}\vspace*{5pt}} 
\setcounter{secnumdepth}{5}

\makeatletter 
\renewcommand\@biblabel[1]{#1}            
\renewcommand\@makefntext[1]% 
{\noindent\makebox[0pt][r]{\@thefnmark\,}#1}
\makeatother 
\renewcommand{\figurename}{\small{Fig.}~}
\sectionfont{\sffamily\Large}
\subsectionfont{\normalsize}
\subsubsectionfont{\bf}
\setstretch{1.125} %In particular, please do not alter this line.
\setlength{\skip\footins}{0.8cm}
\setlength{\footnotesep}{0.25cm}
\setlength{\jot}{10pt}
\titlespacing*{\section}{0pt}{4pt}{4pt}
\titlespacing*{\subsection}{0pt}{15pt}{1pt}
%%%END OF PAGE SETUP%%%

%%%FOOTER%%%
\fancyfoot{}
\fancyfoot[LO,RE]{\vspace{-7.1pt}\includegraphics[height=9pt]{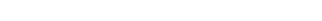}}
\fancyfoot[CO]{\vspace{-7.1pt}\hspace{11.9cm}\includegraphics{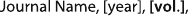}}
\fancyfoot[CE]{\vspace{-7.2pt}\hspace{-13.2cm}\includegraphics{head_foot/RF}}
\fancyfoot[RO]{\footnotesize{\sffamily{1--\pageref{LastPage} ~\textbar  \hspace{2pt}\thepage}}}
\fancyfoot[LE]{\footnotesize{\sffamily{\thepage~\textbar\hspace{4.65cm} 1--\pageref{LastPage}}}}
\fancyhead{}
\renewcommand{\headrulewidth}{0pt} 
\renewcommand{\footrulewidth}{0pt}
\setlength{\arrayrulewidth}{1pt}
\setlength{\columnsep}{6.5mm}
\setlength\bibsep{1pt}
%%%END OF FOOTER%%%

%%%FIGURE SETUP - please do not change any commands within this section%%%
\makeatletter 
\newlength{\figrulesep} 
\setlength{\figrulesep}{0.5\textfloatsep} 

\newcommand{\topfigrule}{\vspace*{-1pt}% 
\noindent{\color{cream}\rule[-\figrulesep]{\columnwidth}{1.5pt}} }

\newcommand{\botfigrule}{\vspace*{-2pt}% 
\noindent{\color{cream}\rule[\figrulesep]{\columnwidth}{1.5pt}} }

\newcommand{\dblfigrule}{\vspace*{-1pt}% 
\noindent{\color{cream}\rule[-\figrulesep]{\textwidth}{1.5pt}} }

\makeatother
%%%END OF FIGURE SETUP%%%

%%%TITLE, AUTHORS AND ABSTRACT%%%
\twocolumn[
  \begin{@twocolumnfalse}\par
\vspace{1em}
\sffamily
\begin{tabular}{m{4.5cm} p{13.5cm} }

 & \noindent\LARGE{\textbf{Log-Gaussian Gamma Processes for Training Bayesian Neural Networks in Raman and CARS Spectroscopies}} \\%Article title goes here instead of the text "This is the title"
\vspace{0.3cm} & \vspace{0.3cm} \\

 & \noindent\large{Teemu Härkönen,$^{\ast}$\textit{$^{a}$} Erik M. Vartiainen,\textit{$^{a}$} Lasse Lensu,\textit{$^{a}$} Matthew T. Moores,\textit{$^{b,a}$} and Lassi Roininen\textit{$^{a}$}} \\%Author names go here instead of "Full name", etc.

 & \noindent\normalsize{%
    We propose an approach utilizing gamma-distributed random variables, coupled with log-Gaussian modeling, to generate synthetic datasets suitable for training neural networks.
    This addresses the challenge of limited real observations in various applications.
    We apply this methodology to both Raman and coherent anti-Stokes Raman scattering (CARS) spectra, using experimental spectra to estimate gamma process parameters.
    %We propose using gamma-distributed random variables with the scale of the distribution modeled as a log-Gaussian process to generate arbitrary quantities of synthetic data for training neural networks in applications where real observations are available in limited quantities.
    %
    %We apply this method to both Raman and coherent anti-Stokes Raman scattering (CARS) spectra, where we use measured spectra to estimate the parameters of the gamma process.
    %
    Parameter estimation  is performed using Markov chain Monte Carlo methods, yielding a full Bayesian posterior distribution for the model which can be sampled for synthetic data generation.
    Additionally, we model the additive and multiplicative background functions for Raman and CARS with Gaussian processes.
    We train two Bayesian neural networks to estimate parameters of the gamma process which can then be used to estimate the underlying Raman spectrum and simultaneously provide uncertainty through the estimation of parameters of a probability distribution.
    We apply the trained Bayesian neural networks to experimental Raman spectra of phthalocyanine blue, aniline black, naphthol red, and red 264 pigments and also to experimental CARS spectra of adenosine phosphate, fructose, glucose, and sucrose.
    The results agree with deterministic point estimates for the underlying Raman and CARS spectral signatures.
    } \\%The abstract goes here instead of the text "The abstract should be..."

\end{tabular}

 \end{@twocolumnfalse} \vspace{0.6cm}

  ]
%%%END OF TITLE, AUTHORS AND ABSTRACT%%%

%%%FONT SETUP - please do not change any commands within this section
\renewcommand*\rmdefault{bch}\normalfont\upshape
\rmfamily
\section*{}
\vspace{-1cm}

%%%FOOTNOTES%%%
\footnotetext{$^{*}$~\textit{Corresponding author, E-mail: teemu.harkonen@lut.fi}}
\footnotetext{\textit{$^{a}$~Department of Computational Engineering, School of Engineering Sciences, LUT University, Yliopistonkatu 34, FI-53850, Lappeenranta, Finland}}
\footnotetext{\textit{$^{b}$~National Institute for Applied Statistics Research Australia, University of Wollongong, Wollongong NSW 2522, Australia}}

%Please use \dag to cite the ESI in the main text of the article.
%If you article does not have ESI please remove the the \dag symbol from the title and the footnotetext below.
%\footnotetext{\dag~Electronic Supplementary Information (ESI) available. See DOI: 10.1039/cXCP00000x/}
%additional addresses can be cited as above using the lower-case letters, c, d, e... If all authors are from the same address, no letter is required

%\footnotetext{\ddag~Additional footnotes to the title and authors can be included \textit{e.g.}\ `Present address:' or `These authors contributed equally to this work' as above using the symbols: \ddag, \textsection, and \P. Please place the appropriate symbol next to the author's name and include a \texttt{\textbackslash footnotetext} entry in the the correct place in the list.}

%%%END OF FOOTNOTES%%%

%%%%%%%%%%%%%%%%%%%%%%%%%%%%%%%%%%%%%%%%
\section{Introduction}
\label{sec:introduction}
%
%Both Raman and coherent anti-Stokes  Raman scattering (CARS) spectroscopy provide tools to investigate molecular samples in a plethora of fields including medicine and physics\cite{Mulvaney:2000, Krafft:2012, Day:2011}.
%
Raman and coherent anti-Stokes Raman scattering (CARS) spectroscopies are vital tools used in chemistry, physics, and biomedical research\cite{Mulvaney:2000, Krafft:2012, Day:2011}.
The insights they offer into molecular vibrations, structural dynamics, and chemical compositions are invaluable.
However, working with their data presents challenges.
%
%Extensive descriptions of the underlying physical interactions of Raman and CARS spectroscopy can be found in a variety of sources and we refer the reader to see for example in the references cited above.
%
%In short, both Raman and CARS spectra result by inelastic photon scattering where the incident photons interact with the molecular vibrations and, thus, the detected scattered photons provide information on the vibrational energy structure of the sample molecules.
%In short, both Raman and CARS spectroscopy excite molecular bonds to higher energy level via lasers.
%
%When the energy level shifts back to its lower energy state, a photon is emitted.
%
%The emitted photons are then measured, constituting the measurement spectrum.
%
%The measurement spectra are contaminated with measurement noise and other undesirable artefacts, namely either with an additive or multiplicative background signal.
%
Measurement artifacts including noise, and, especially, background signals in Raman and CARS spectra often obscure crucial molecular information.
%
%
%The removal of the measurement noise and the background signals yields the underlying Raman spectrum which holds the chemically specific information about the measurement sample.
%
Traditional methods for data correction are typically manual and may fall short in capturing the full complexity of the data.
For instance, standard approaches used for removing the background signals include asymmetric least squares polynomial fitting, wavelet-based methods, optimization with Tikhonov regularization, and Kramers-Kronig relations \cite{Boelens:2005, He:2014, Gan:2006, Galloway:09, Vartiainen:06, Kan:16, Chi:2019, Harkonen:2023, Liland:2010}.
While appealing, these methods suffer from practical drawbacks such as the need for manual tuning of the model or regularization parameters.
The need for automated, robust, and statistically sound solutions to enhance our spectroscopic analyses is evident.

%Deep neural networks can provide a completely automatic approach for correcting spectral measurements amongst their extensive other application areas\cite{Sharma:2017, Shinde:2018, Samek:2021, Alzubaidi:2021, Li:2022}.
%
Deep neural networks offer a compelling solution for automatic spectral correction across various applications, from weather predictions \cite{ Weyn:2020, Ritvanen:2023, Abdalla:2021} to medical imaging\cite{Hamilton:2018, Monti:2020, Suganyadevi:2022} and many others \cite{Sharma:2017, Shinde:2018, Samek:2021, Alzubaidi:2021, Li:2022}.
In the realm of Raman spectroscopy, deep neural networks have been used in chemical species identification and background removal \cite{Liu:2017, Wahl:2020, Gebrekidan:2021, Kazemzadeh:2022, Luo:2022}.
Similarly, they have been applied to extract the underlying Raman spectra from CARS measurement \cite{Valensise:2020, Houhou:2020, Wang:2022, Junjuri:2022, Saghi:2022, Junjuri:2022b, Junjuri:2023}.
Despite their efficacy, non-Bayesian neural networks lack a critical feature: the ability to quantify uncertainty in Raman spectrum estimation.
Bayesian inference, on the other hand, provides an avenue to solve this problem.

Bayesian inference treats the parameters of a given model as random variables.
These models consist of a likelihood function that is combined with prior distributions for the parameters to produce posterior estimates.
The likelihood function is analogous to a utility function in an optimization context. It quantifies how well the model fits the observed data.
The aforementioned prior distributions for the model parameters represent the information known beforehand, including any constraints dictated by the physical nature of the parameters, such as non-negativity.
In spectroscopic analysis, the model parameters can be, for example, amplitudes, locations, and widths of Gaussian, Lorentzian, or Voigt line shape functions.
The combination of the likelihood and the priors  results in a posterior  distribution over the model parameters.
The posterior is a probabilistic representation of the uncertainty in the parameter estimates.
Bayesian approaches have been considered for estimating spectrum parameters, where the authors used sequential Monte Carlo algorithms to numerically sample from the posterior distribution~\cite{Moores:16, Harkonen:2020}.
While the uncertainty quantification provided by Bayesian modeling and Markov chain Monte Carlo (MCMC) methods is compelling, the approach is known to be computationally expensive, see for example~\cite{Wang:2015}.
This becomes a major issue particularly with hyperspectral data sets.
A hyperspectral data set, or an image, consists of pixels where each pixel contains a spectrum.
This can quickly result in millions of individual spectra, experimental or synthetic, which are to be analyzed.

Bayesian neural networks are a synthesis of the aforementioned two ideas.
Bayesian neural networks model the weights and biases of standard neural networks as random variables, which can be assigned prior distributions.
When combined with a likelihood according to Bayes' theorem, the resulting utility function corresponds to the posterior for the neural network parameters.
Advantages of this Bayesian neural network approach in comparison to non-Bayesian neural networks include robustness in terms of overfitting, providing uncertainty estimates instead of only point estimation, sequential learning, and better generalization~\cite{Magris:2023}.
In particular, uncertainty quantification has seen widespread research covering many application areas and topics, for example~\cite{Abdar:2021}.

One of the challenges of Bayesian neural networks is that they typically contain an enormous number of parameters.
For instance, our network comprises over 11 million parameters, far beyond what is commonly considered high-dimensional for MCMC~\cite{Cui:2016, Morzfeld:2019}.
Some neural networks, such as large language models (LLMs), can have billions of parameters~\cite{Yang:2023}.
Thus, it can be challenging to establish convergence of such a large number of parameters in a statistically rigorous manner.
To combat this, partially-Bayesian neural networks have been used as a practical tool to provide uncertainty estimation with neural networks.
In addition to empirical validation through practice, studies have provided  compelling analytical and numerical evidence that partially-Bayesian neural networks are indeed capable of providing posterior estimates on par or even superior performance to fully-Bayesian neural networks~\cite{Sharma:2023}.
The above points lead us to construct our neural network for this study as a partially-Bayesian neural network.

\begin{figure}[!ht]
    \centering
    \includegraphics[width = 0.5\textwidth]{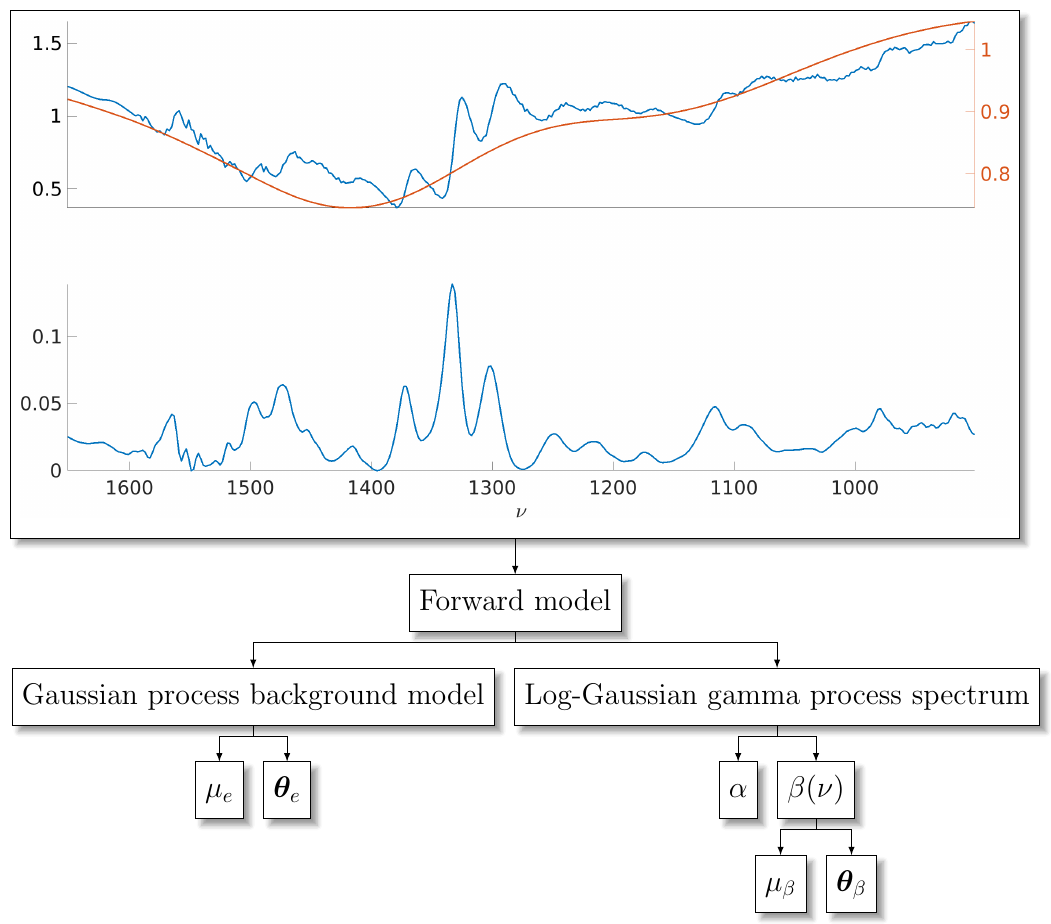}
    \caption{Structure of our generative spectrum model using GPs and log-Gaussian gamma processes. On top, an experimental CARS spectrum of adenosine phosphate in blue and an example multiplicative background in red. We model the backgrounds as a GP. At the bottom, an example underlying Raman spectral signature in blue. We assume the Raman peaks to be distributed according to our proposed log-Gaussian gamma process model. The stochastic processes are parameterized according to $\mu_e$, $\bm\theta_e$, $\alpha$, and $\beta(\nu)$. We further model $\beta(\nu)$ using GPs which are parameterized according to $\mu_\beta$ and $\bm\theta_\beta$. We construct statistical samples with MCMC for the model parameters which allow us to generate synthetic spectra for training our Bayesian neural network.}
    \label{im:spectrumWorkflow}
\end{figure}

\begin{figure}[!htpb]
    \centering
    \includegraphics[width = 0.5\textwidth]{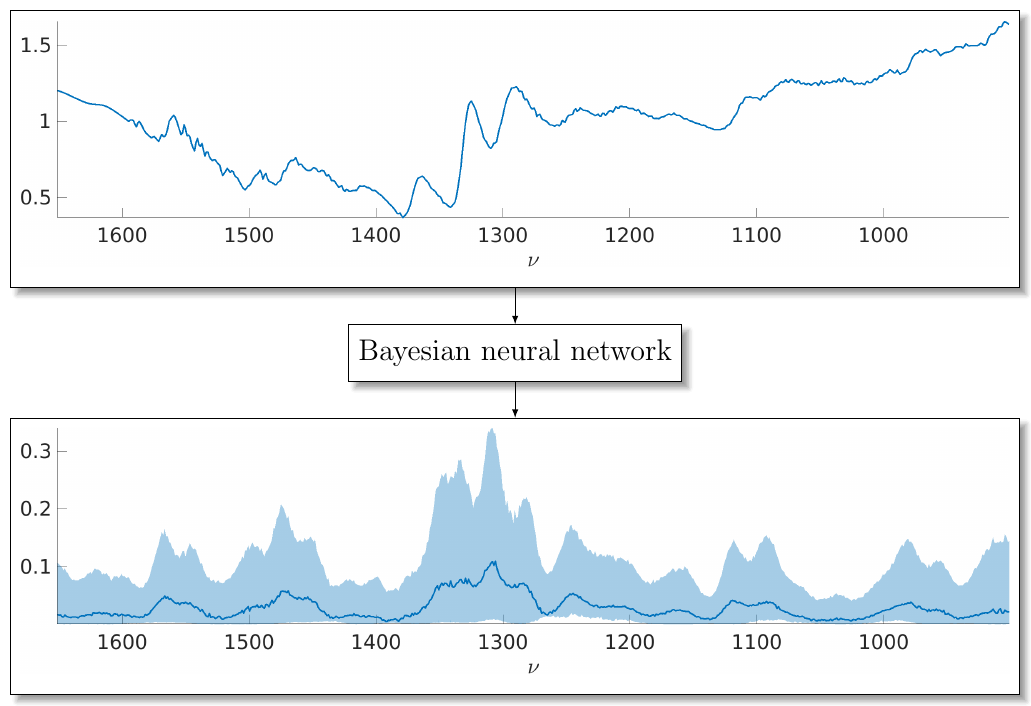}
    \caption{On top, an experimental CARS spectrum of adenosine phosphate in blue. With a trained Bayesian neural network, we can extract the underlying Raman spectral signature from the data along with an uncertainty estimate for the spectrum. At the bottom, the corresponding Bayesian neural network median Raman spectrum estimate and 90\% confidence interval of the estimate for the adenosine phosphate data.}
    \label{im:bnnExample}
\end{figure}

Neural networks typically require large volumes of training data.
This has been noted to be a problem also in spectroscopic applications as it is difficult to acquire large sets of independent data sets~\cite{Luo:2022}.
Therefore, many studies mentioned above use synthetic data to train the neural networks.
The synthetic data is usually generated using random linear combinations of Lorentzian line shapes, where the  amplitudes, locations, and widths are sampled from predefined probability distributions, see for example~\cite{Valensise:2020, Wahl:2020, Antonio:2023}.
The background data is generated similarly.
The backgrounds are modeled explicitly using a parametric functional form, such as a polynomial or a sigmoidal function, and the parameters of the model are again sampled from a predefined probability distribution~\cite{Wahl:2020, Mozaffari:2021, Junjuri:2022}.
An extension to this is to use experimental Raman spectra on top of the randomly generated spectra~\cite{Junjuri:2022b}.
Stochastic processes can be used to draw samples of random functions.
A typical example of a stochastic process is the widely-used Gaussian process (GP).
Properties of the drawn samples such as differentiability are governed through kernel functions, which are used to model dependencies between data points.
For readers unfamiliar with GPs, we recommend the book by Rasmussen and Williams~\cite{Rasmussen:2005}.
Instead of using explicit, parametric functions to model the spectroscopic features, we propose using stochastic processes as a more flexible tool for the purpose.
In this study, we use GPs as a generative model for the additive and multiplicative backgrounds of Raman and CARS spectra, see Fig.~\ref{im:spectrumWorkflow}.
For the purpose of generating synthetic Raman spectral signatures, we propose a specific type of doubly-stochastic L{\'e}vy process which we call a log-Gaussian gamma process.
Our construction of the log-Gaussian gamma process is inspired by log-Gaussian Cox process which the authors have previously used as a model for spectra~\cite{HarkonenFoDS:2023}.
While it makes sense to model spectra as a Cox process where the relaxation from higher energy levels happens at a constant rate and results in counts of photons, the data is often available in scaled floating-point numbers which prevents direct application of the log-Gaussian Cox process model.
Gamma-distributed variables have direct connections to Poisson-distributed variables, which constitute the Cox process, making the extension to a log-Gaussian gamma process intuitive as a model for Raman spectroscopy.
The log-Gaussian gamma process can be used to generate arbitrary amounts of synthetic spectra once parameters of the stochastic process have been estimated.
We perform the estimation using MCMC methods which allow us to construct a Bayesian posterior distribution for the model parameters, thereby including uncertainty of the parameter estimates in our data generation.
This also applies to our GP-based background model.
We present a high-level diagram of our stochastic process method for data generation in Fig.~\ref{im:spectrumWorkflow}.
Fig.~\ref{im:bnnExample} shows an example of the aim of this paper, a Raman spectral signature extracted from a CARS spectrum using a Bayesian neural network.
We provide a pseudo-code description of our approach in Algorithm~\ref{alg:lggpBNN}.
The key contributions of this paper are the following.
We propose using log-Gaussian gamma processes for modeling Raman spectral signatures and GPs to model additive or multiplicative background signals.
The aforementioned doubly-stochastic processes are sampled randomly,  enabling us to generate an arbitrary number of synthetic spectra that are statistically similar to experimental spectra.
Finally, we present a partially-Bayesian neural network for analyzing Raman and CARS spectral measurements, which we train using the sampled synthetic spectra. 
Once trained, we use these neural networks to estimate the spectral signatures for experimental Raman spectroscopy measurements of phthalocyanine blue, naphthol red, aniline black, and red 264 pigments and for experimental CARS spectra of adenosine phosphate, fructose, glucose, and sucrose in addition to synthetic test spectra.
\begin{algorithm}
    \caption{Log-Gaussian gamma process data generation for training Bayesian neural networks}
    \label{alg:lggpBNN}
    \begin{algorithmic}
        \State Step 1: Fit a log-Gaussian gamma process to Raman spectrum data.
        \State Step 2: Fit a GP to background data.
        \State Step 3: Draw a large number of realizations from the fitted log-Gaussian gamma process.
        \State Step 4: Draw a large number of realizations from the fitted GP.
        \State Step 5: Use a forward model to combine the realizations to form a data set of synthetic spectra.
        \State Step 6. Train a Bayesian neural network using the data set of synthetic spectra.
    \end{algorithmic}
\end{algorithm}
The rest of the paper is structured as follows.
We detail the steps used to generate the synthetic training data in three stages in the following three sections.
We first present the log-Gaussian gamma process as a model for Raman spectral signatures and explain how to draw realizations of this doubly-stochastic process.
This is followed by a description of our GP-based additive and multiplicative background models.
We finalize the explanation of our synthetic data generation method with definitions of the forward models used to simulate synthetic training data for Raman and CARS measurements with additive and multiplicative backgrounds, respectively.
Next, we present our partially-Bayesian neural network architecture, which we train against the synthetic data sets that we have generated.
We document computational details and prior distributions in the next section, followed by a presentation of our results for both artificial and real experimental data.
Finally, we conclude with a discussion of the significance and other potential applications for our method.
%
%%%%%%%%%%%%%%%%%%%%%%%%%%%%%%%%%%%%%%%%
\section{Log-Gaussian gamma process spectrum model}
\label{sec:lggp_model}
We model a Raman spectral signature as a collection of conditionally-independent, gamma-distributed random variables
\begin{equation}
    r_k := r( \nu_k ) \sim \rm{Gamma}\left( \alpha, \beta( \nu_k ) \right),
    \label{eq:gammaSpectrumDistribution}
\end{equation}
where $r_k$ denotes a Raman measurement at wavenumber location $\nu_k$ with $\alpha$ and $\beta( \nu_k )$ being the shape and scale parameters of the gamma distribution, respectively.
The scale $\beta( \nu_k )$ is thought to model the spectral line shapes and other artefacts, while the shape $\alpha$ models the noise level present in the spectrum.
The above construction is motivated by log-Gaussian Cox processes\cite{Moller:1998} but without the restriction of modeling of only integer-valued data and with an additional parameter in the stochastic process allowing for more flexible modeling of uncertainty.
Poisson-distributed random variables, which constitute the Cox process, have a single parameter to control both the mean and variance of the distribution.
Very often in real data, this assumption is found to be too restrictive, leading to a model that is either under- or over-dispersed \cite{Hilbe:2011}.
In contrast, the gamma distribution has two parameters which together allow for a range of different variances for a given mean.

We extend Eq.~\eqref{eq:gammaSpectrumDistribution} by modeling the log-scale as a GP, resulting in a hierarchical model
\begin{equation}
    \log\beta( \bm\nu ) \sim \rm{GP}( \mu_{\beta}, \Sigma_\beta( \bm\nu, \bm\nu, \bm\theta_\beta)),
\end{equation}
where $\bm\nu := (\nu_1, \dots, \nu_K)^\intercal$  is a vector of the wavenumber locations with $\mu_{\beta} \in \mathbb{R}$ and $ \Sigma_\beta( \bm\nu, \bm\nu, \bm\theta_\beta) \in \mathbb{R}^{K \times K}$ being a constant mean and a covariance matrix parameterized according to hyperparameters $\bm\theta_\beta$.
This doubly-stochastic model introduces dependence between values $r_i$ and $r_j$ at different wavenumbers $\nu_i$ and $\nu_j$.
For the covariance function of the log-scale GP, we use the squared exponential kernel
\begin{equation}
    \left[ \Sigma_\beta( \bm\nu, \bm\nu, \bm\theta_\beta) \right]_{ij} = \sigma_{\beta, f}^2 \exp\left( -\frac{1}{2} \frac{\left(\nu_i - \nu_j\right)^2}{l_\beta^2} \right) + \sigma_{\beta}^2 \delta(\nu_i - \nu_j),
    \label{eq:errorf_covariance_beta}
\end{equation}
where $\left[ \Sigma_e( \bm\nu, \bm\nu, \bm\theta_\beta) \right]_{ij}$ denotes the $ij$th element of the covariance matrix $\Sigma_\beta( \bm\nu, \bm\nu, \bm\theta_\beta)$, $\sigma_{\beta, f}^2$ is the signal variance, $l_\beta$ is the length scale, $\sigma_{\beta}^2$ denotes the noise variance, and $\delta(\nu_i - \nu_j)$ is the Dirac delta function with $\bm\theta_\beta = \left( \sigma_{\beta, f}^2, l_\beta, \sigma_{\beta}^2\right)^\intercal$.
The GP construction yields an analytical form for the log-scale $\log\beta( \bm\nu )$ which we will detail below as we construct the posterior distribution according to Bayes' theorem.
This log-GP parameterization is identical to the log-intensity model for Poisson variables that features in log-Gaussian Cox processes.
For more details on the log-Gaussian Cox process, see \cite{Moller:1998} and for example \cite{Teng:2017}.

The posterior distribution involves the  likelihood function $\mathcal{L}( \bm{r} \mid \alpha, \beta( \bm\nu ) )$, a log-GP prior for the scale $\pi_0( \beta(\bm\nu) \mid \mu_\beta, \bm\theta_\beta)$, and a joint prior distribution $\pi_0( \alpha, \mu_\beta, \bm\theta_\beta)$ for rest of the model parameters.
Given a measured Raman spectrum $ \bm{r} := ( r( \nu_1 ), \dots, r( \nu_K) )^\intercal$, we can formulate the likelihood as a product of conditionally-independent, gamma-distributed random variables
\begin{equation}
    \mathcal{L}( \bm{r} \mid \alpha, \beta( \bm\nu ) ) \propto \prod\limits_{k = 1}^K \frac{ r_k^{\alpha - 1}\exp( -r_k/\beta_k ) }{ \Gamma(\alpha) \beta_k^\alpha },
\end{equation}
where $\beta_k := \beta(\nu_k)$, and $\Gamma( \alpha )$ is the gamma function.
The hierarchical prior for $\beta(\bm\nu)$ can be evaluated as
\begin{equation}
\begin{split}
     \pi_0( &\beta(\bm\nu) \mid \mu_\beta, \bm\theta_\beta) = \frac{1}{\sqrt{(2\pi)^K}} \left\vert \Sigma_\beta( \bm\nu, \bm\nu; \bm\theta_\beta) \right\vert^{-1/2}\\
     &\times \exp\left(
     -\frac{1}{2} \left( \beta(\bm\nu) - \mu_\beta \right)^\intercal \Sigma_\beta( \bm\nu, \bm\nu; \bm\theta_\beta)^{-1} \left( \beta(\bm\nu) - \mu_\beta \right) \right),
\end{split}
    \label{eq:beta_gp_loglikelihood}
\end{equation}
where $\left\vert \Sigma_\beta( \bm\nu, \bm\nu; \bm\theta_\beta) \right\vert$ denotes the determinant of the covariance matrix.
With the above and a joint prior $\pi_0( \alpha, \mu_\beta, \bm\theta_\beta)$, we can construct the posterior distribution for the model parameters conditioned on the measured spectrum data $ \bm{r} $ as
\begin{equation}
\begin{split}
    \pi( \alpha, \beta( \bm\nu ), \mu_\beta, \bm\theta_\beta \mid \bm{r}) \propto \mathcal{L}( \bm{r} \mid \alpha, \beta( \bm\nu )) &\pi_0( \beta(\bm\nu) \mid \mu_\beta, \bm\theta_\beta)\\ &\times\pi_0( \alpha, \mu_\beta, \bm\theta_\beta).
\end{split}
    \label{eq:lggp_posterior}
\end{equation}
In the posterior in Eq.\eqref{eq:lggp_posterior}, the dimension of $\beta( \bm\nu )$ is $K$.

The scale is a vector of the same dimension as the data, $\beta( \bm\nu ) \in \mathbb{R}_+^{K \times 1}$.
MCMC methdos are known to struggle estimating high-dimensional parameters.
At a minimum, the high-dimensional parameters incur a computational cost for inference with MCMC.
To amend these issues and to simplify the inference, we perform dimension reduction for the scale $\beta( \bm\nu )$.
To achieve this, we observe that our data $\bm{r}$ should be a reasonable estimate for the expectation of the gamma process in Eq.\eqref{eq:gammaSpectrumDistribution}, $\bm{r} \approx \mathbb{E}[ \rm{Gamma}( \alpha, \beta( \bm\nu) ) ] = \alpha \beta( \bm\nu)$.
This implies that the shape of the data $\bm{r}$ is close to the shape of the scale function, $\beta( \bm\nu)$.
Thus, we approximate the scale $\beta( \bm\nu)$ as a convolution between a Gaussian kernel and the data 
\begin{equation}
    \beta( \bm\nu ) \approx c_\beta G( \bm\nu; \sigma_G) \ast \bm{r},
\end{equation}
where $\ast$ denotes convolution, $c_\beta$ is a scaling constant, and $G(\bm\nu; \sigma_G)$ is Gaussian smoothing kernel with width $\sigma_G$.
By this, we reduce the inference of the scale $\beta( \bm\nu ) \in \mathbb{R}_+^{K \times 1}$ to inference of two parameters, $c_\beta$ and $\sigma_G$.
With this smoothing approximation, we formulate an approximate posterior for Eq.~\eqref{eq:lggp_posterior} as
\begin{equation}
\begin{split}
    \pi( \alpha, c_\beta, \sigma_G, \mu_\beta, \bm\theta_\beta \mid \bm{r}) = \widetilde{\mathcal{L}}( \bm{r} \mid \alpha, c_\beta, &\sigma_G) \pi_0( \beta(\bm\nu) \mid \mu_\beta, \bm\theta_\beta)\\ &\times\pi_0( \alpha, c_\beta, \sigma_G, \mu_\beta, \bm\theta_\beta),
\end{split}
    \label{eq:approximate_lggp_posterior}
\end{equation}
where $\widetilde{\mathcal{L}}( \bm{r} \mid \alpha, c_\beta, \sigma_G) = \mathcal{L}( \bm{r} \mid \alpha, c_\beta G( \bm\nu; \sigma_G) \ast \bm{r} )$ and $\pi_0( \alpha, c_\beta, \sigma_G, \mu_\beta, \bm\theta_\beta)$ is the prior distribution augmented with $(c_\beta, \sigma_G)^\intercal$.
We detail the prior distribution $\pi_0( \alpha, c_\beta, \sigma_G, \mu_\beta, \bm\theta_\beta)$ in the section on computational details and prior distributions.
We perform inference of the posterior in Eq.~\eqref{eq:approximate_lggp_posterior} by sampling all the model parameters simultaneously using the DRAM algorithm \cite{Haario:2006}.

Given samples from the posterior distribution $\pi( \alpha, c_\beta, \sigma_G, \mu_\beta, \bm\theta_\beta \mid \bm{r})$ obtained with MCMC, we can sample realizations for the synthetic spectra to generate an arbitrary amount of synthetic data in the following way.
First, we sample the GP parameters $( \widetilde{\mu}_\beta, \widetilde{\bm\theta}_\beta)^\intercal$ from the MCMC chain.
Next, we use $( \widetilde{\mu}_\beta, \widetilde{\bm\theta}_\beta)^\intercal$ sample a GP realization $\widetilde{\beta}( \bm\nu^* \mid \widetilde{\mu}_\beta, \widetilde{\bm\theta}_\beta )$ at prediction locations $ \bm\nu^* := ( \nu_1^*, \dots, \nu_{\widetilde{K}})^\intercal$ modeling the scale $\beta( \bm\nu^* )$ with
\begin{equation}
      \widetilde{\beta}( \bm\nu^* \mid \widetilde{\mu}_\beta, \widetilde{\bm\theta}_\beta ) = \exp\left( \widetilde{\mu}_\beta + L(\bm\nu^* \mid \widetilde{\bm\theta}_\beta)\bm{u} \right),
\end{equation}
where $L(\bm\nu^* \mid \widetilde{\bm\theta}_\beta)$ is the lower triangular Cholesky decomposition matrix of $\Sigma_\beta( \bm\nu^*, \bm\nu^*; \widetilde{\bm\theta}_\beta)$ and $\bm{u} := ( u_1, \dots, u_{\widetilde{K}} )^\intercal$ is Gaussian white noise such that $ u_{\widetilde{k}} \sim \mathcal{N}( 0, 1)$.
Finally, by sampling $\widetilde{\alpha}$, we can draw a spectrum realization $\widetilde{r}(\bm\nu)$ from the gamma process,  $ \rm{Gamma}( \widetilde{\alpha}, \widetilde{\beta}( \bm\nu) )$.

\begin{figure}
    \centering
    \includegraphics[width = 0.5\textwidth]{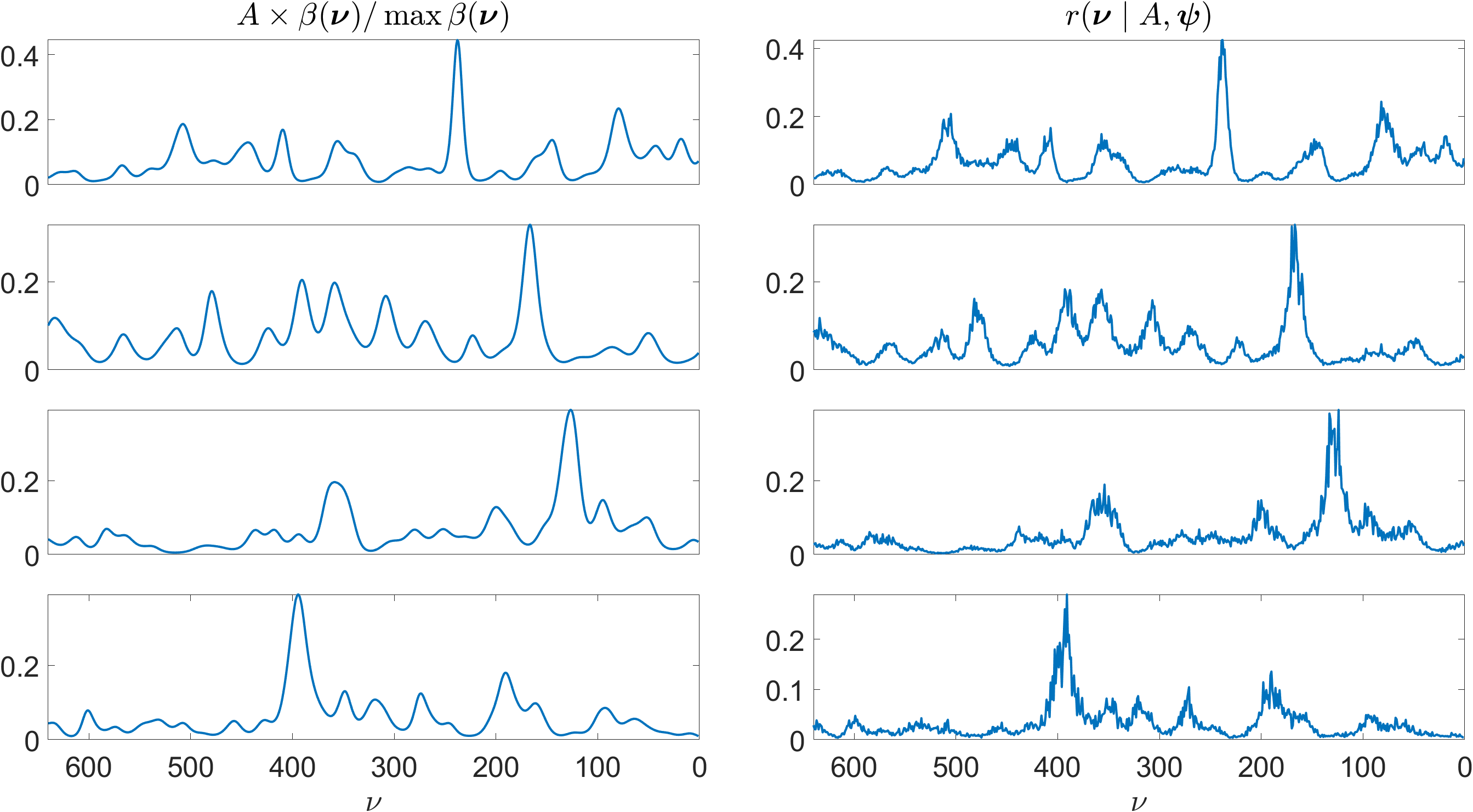}
    \caption{Example realizations drawn from the log-Gaussian gamma process model defined in Eq.~\eqref{eq:lggp_statistical_spectrum_model}. On the left, realizations for the scale process $\beta(\bm\nu)$, drawn from a log-Gaussian process. On the right, corresponding realizations from the gamma process. All realizations are normalized and multiplied by a sampled amplitude.}
    \label{im:exampleRealizationsLGGP}
\end{figure}

We normalize the realizations $\widetilde{r}(\bm\nu)$ such that $ \max \left\{ \widetilde{r}(\bm\nu) \right\} = 1$ and introduce an additional parameter to control amplitudes of the realizations.
With an amplitude parameter $A$, we sample a normalized shape $\widetilde{r}_N(\bm\nu \mid \bm\psi )$ of the spectrum and multiply this by a sampled amplitude $\widetilde{A}$.
This procedure results in the following statistical model
\begin{equation}
    \begin{split}
        r(\bm\nu \mid A, \bm\psi) &\sim A \frac{r_N(\bm\nu \mid \bm\psi )}{\max r_N(\bm\nu \mid \bm\psi )}, \\
        r_N(\bm\nu \mid \bm\psi ) &\sim \rm{Gamma}( \alpha, \beta( \bm\nu) ),\\
        A &\sim \pi_0( A ),\\
        \bm\psi &\sim \pi_0( \bm\psi ),
    \end{split}
    \label{eq:lggp_statistical_spectrum_model}
\end{equation}
where $\bm\psi := (\alpha, c_\beta, \sigma_G, \mu_\beta, \bm\theta_\beta)^\intercal$ is a shorthand for the gamma process parameters and $\pi_0(A)$ is a prior distribution for the amplitude $A$.
Example realizations from the above statistical model are shown in Fig. \ref{im:exampleRealizationsLGGP}.
In the following section, we detail how we model additive and multiplicative backgrounds for Raman and CARS spectra using GPs.
%
%
%%%%%%%%%%%%%%%%%%%%%%%%%%%%%%%%%%%%%%%%
\section{Additive and multiplicative background models}
\label{sec:error_model}
We propose GPs as a flexible way to randomly draw additive and multiplicative background functions for Raman and CARS spectrum modeling.
This is in contrast to more standard polynomial models such as the ones used in \cite{Junjuri:2022}.

As noted above, we model additive or multiplicative spectral backgrounds as a GP  
\begin{equation}
    e(\bm\nu) \sim \rm{GP}( \mu_e, \Sigma_e( \bm\nu, \bm\nu, \bm\theta_e )),
\end{equation}
with $\mu_e \in \mathbb{R}$ and $ \Sigma_e( \bm\nu, \bm\nu, \bm\theta_e) \in \mathbb{R}^{K \times K}$ being a constant mean and the covariance matrix of the GP parameterized according to hyperparameters $\bm\theta_e$.
For the background GP covariance function, we use again the squared exponential kernel
\begin{equation}
    \left[ \Sigma_e( \bm\nu, \bm\nu, \bm\theta_e) \right]_{i,j} = \sigma_{ e,f}^2 \exp\left( -\frac{1}{2} \frac{\left(\nu_i - \nu_j\right)^2}{l_e^2} \right) + \sigma_e^2 \delta(\nu_i - \nu_j)
    \label{eq:errorf_covariance}
\end{equation}
where $\left[ \Sigma_e( \bm\nu, \bm\nu, \bm\theta_e) \right]_{i,j}$ denotes the $ij$th element of the covariance matrix, $\sigma_{ e, f}^2$ is the signal variance, $l_e$ is the length scale, and $\sigma_e^2$ denotes the noise variance with $\bm\theta_e := \left( \sigma_{ e,f}, l_e, \sigma_e \right)^\intercal$.

Given a measurement of the background process, $\bm{e} := ( e(\nu_1), \dots, e(\nu_K))^\intercal$, we can formulate a posterior distribution for the background GP parameters $(\mu_e, \bm\theta_e)^\intercal$ as
\begin{equation}
    \pi( \mu_e, \bm\theta_e \mid \bm{e}) \propto \mathcal{L}( \bm{e} \mid \mu_e, \bm\theta_e ) \pi_0( \mu_e, \bm\theta_e ),
    \label{eq:background_posterior}
\end{equation}
where $\mathcal{L}( \bm{e} \mid \mu_e, \bm\theta_e)$ is the GP likelihood and $\pi_0( \mu_e, \bm\theta_e )$ denotes the prior distribution for the GP parameters.
The log-likelihood
%$ \log \mathcal{L}( \bm{e} \mid \mu_e, \bm\theta_e)$
is given as
\begin{equation}
\begin{split}
    \log \mathcal{L}( \bm{e} \mid \mu_e, \bm\theta_e) = &-\frac{1}{2} ( \bm{e} - \mu_e)^\intercal \Sigma_e( \bm\nu, \bm\nu, \bm\theta_e)^{-1} ( \bm{e} - \mu_e) \\&-\frac{1}{2} \log \left\vert \Sigma_e( \bm\nu, \bm\nu, \bm\theta_e) \right\vert- \frac{K}{2} \log 2\pi,
\end{split}
\end{equation}
where $\left\vert \Sigma_e( \bm\nu, \bm\nu, \bm\theta_e) \right\vert$ is the determinant of the covariance matrix.
Again, we perform the posterior estimation for Eq.~\eqref{eq:background_posterior} by sampling all the model parameters simultaneously using DRAM.

\begin{figure}
    \centering
    \includegraphics[width = 0.5\textwidth]{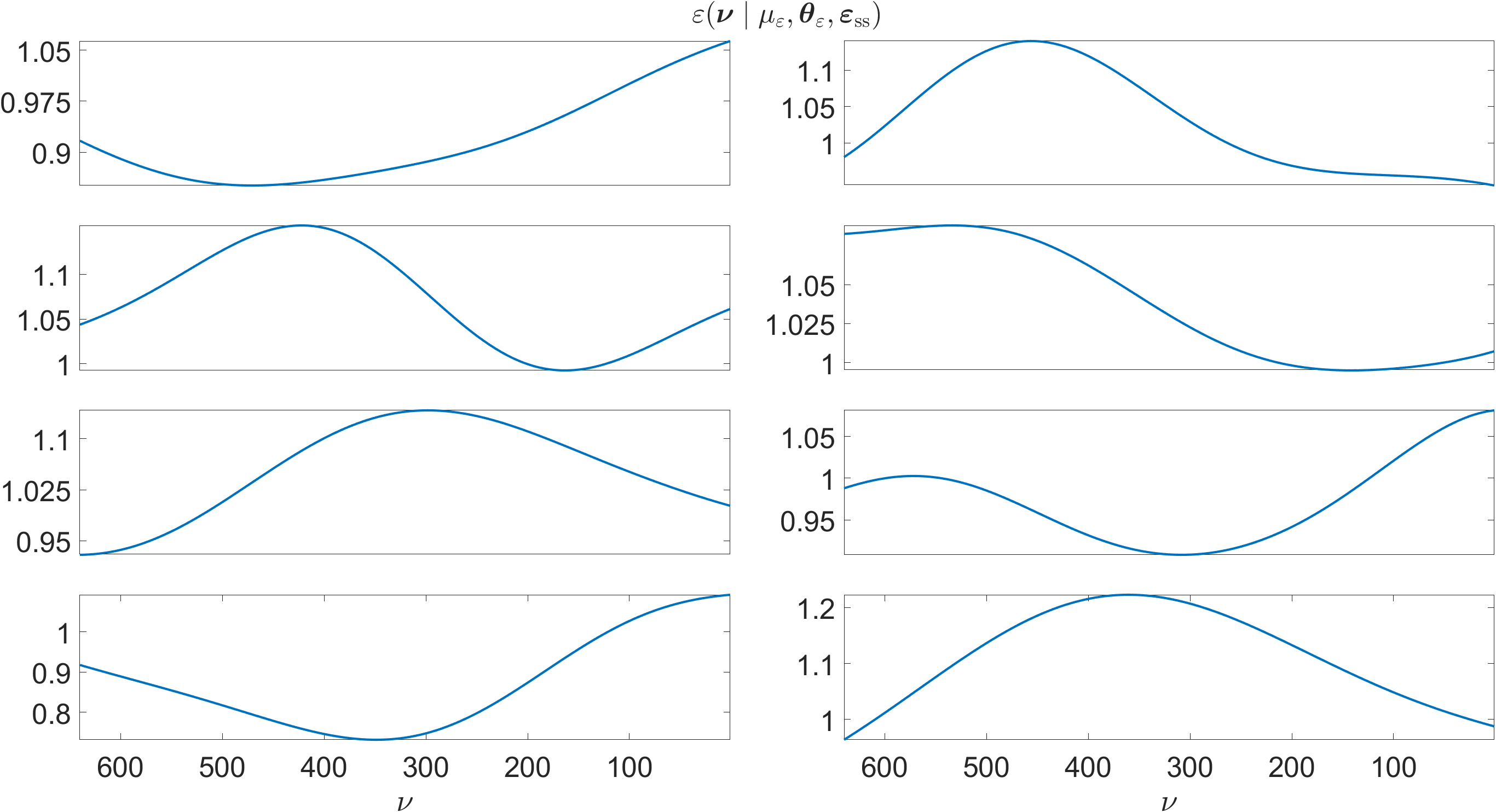}
    \caption{Example realizations drawn from the background model defined in Eq.~\eqref{eq:background_statistical_model} for a multiplicative background. The starting and end points are sampled from a prior distribution and the GP predictive mean and covariance are used to sample the background shape.}
    \label{im:exampleRealizationsBackground}
\end{figure}

Given a posterior $\pi( \mu_e, \bm\theta_e \mid \bm{e})$, we construct realizations for the spectrum by drawing realizations from the GP predictive distribution.
We sample starting and ending points for the background function from priors $\pi_0( e_{\rm start} )$ and $\pi_0( e_{\rm stop} )$, $\pi_0( e_{\rm start}, e_{\rm stop}) = \pi_0( e_{\rm start} )\pi_0( e_{\rm stop} )$.
Next, we compute the predictive mean
\begin{equation}
\begin{split}
    e^*(\bm\nu \mid \widetilde{\mu}_\varepsilon, \widetilde{\bm\theta}_e, \widetilde{\bm\varepsilon}_{\rm ss}) = \Sigma_e( \bm\nu, \bm\nu_{\rm ss}; \widetilde{\bm\theta}_e) \Sigma_e( \bm\nu_{\rm ss}, \bm\nu_{\rm ss}; \widetilde{\bm\theta}_e)^{-1}( \widetilde{\bm e}_{\rm ss} - \widetilde{\mu}_\varepsilon) + \widetilde{\mu}_\varepsilon,
\end{split}
    \label{eq:predictiveMean}
\end{equation}
and the predictive covariance
\begin{equation}
\begin{split}
    \Sigma_e^*( \bm\nu, \bm\nu; \widetilde{\bm\theta}_e) = \Sigma_e( \bm\nu, \bm\nu; \widetilde{\bm\theta}_e) - \Sigma_e( \bm\nu, \bm\nu_{\rm ss}; \widetilde{\bm\theta}_e) &\Sigma_e( \bm\nu_{\rm ss}, \bm\nu_{\rm ss}; \widetilde{\bm\theta}_e) ^{-1}\\ &\times\Sigma_e( \bm\nu, \bm\nu_{\rm ss}; \widetilde{\bm\theta}_e)^\intercal,
\end{split}
    \label{eq:predictiveCovariance}
\end{equation}
where $(\widetilde{\mu}_\varepsilon, \widetilde{\bm\theta}_e)$ are samples from the posterior distribution $\pi( \mu_e, \bm\theta_e \mid \bm{e})$ obtained via MCMC, and $\bm\nu_{\rm ss} = (\nu_{\rm start}, \nu_{\rm stop})^\intercal$ are the wavenumber locations corresponding to the sampled starting and ending points $\widetilde{\bm e}_{\rm ss} = ( \widetilde{e}_{\rm start}, \widetilde{e}_{\rm stop})^\intercal$.
Elements of the covariance matrix $\Sigma_e( \bm\nu, \bm\nu; \widetilde{\bm\theta}_e)$ are given as defined in Eq.~\eqref{eq:errorf_covariance} and elements of the covariance matrices $\Sigma( \bm\nu, \bm\nu_{\rm ss}; \widetilde{\bm\theta}_e)$ and $\Sigma( \bm\nu_{\rm ss}, \bm\nu_{\rm ss}; \widetilde{\bm\theta}_e)$ are given by otherwise the same covariance function but without the diagonal elements produced by the Dirac delta function.
With the above mathematical machinations, we can sample realizations for the background function by
\begin{equation}
    \widetilde{e}(\bm\nu \mid \widetilde{\mu}_e, \widetilde{\bm\theta}_e, \widetilde{\bm e}_{\rm ss}) \sim e^*(\bm\nu \mid \widetilde{\mu}_e, \widetilde{\bm\theta}_e,\widetilde{\bm e}_{\rm ss}) + L(\bm\nu \mid \widetilde{\mu}_e, \widetilde{\bm\theta}_e)\bm{w},
\end{equation}
where $L(\bm\nu \mid \widetilde{\mu}_e, \widetilde{\bm\theta}_e)$ is the lower triangular Cholesky decomposition matrix of $\Sigma^*( \bm\nu, \bm\nu; \widetilde{\bm\theta}_e)$ and $\bm{w} \in \mathbb{R}^{K \times 1}$ is a Gaussian white noise vector.
This is compiled into the following statistical model for the background function sampling:
\begin{equation}
\begin{split}
    \widetilde{e}(\bm\nu \mid \mu_e, \bm\theta_e, \bm{e}_{\rm ss}) &\sim \mathcal{N}( \varepsilon^*, \Sigma^*),\\
    ( \mu_e, \bm\theta_e, \bm{e}_{\rm ss} ) &\sim \pi( \mu_e, \bm\theta_e \mid \bm{e}) \pi_0( \bm e_{\rm ss} ),
\end{split}
\label{eq:background_statistical_model}
\end{equation}
where $\pi_0( \bm{e}_{\rm ss} ) = \pi_0( e_{\rm start}, e_{\rm stop})$.
Example realizations for a multiplicative background relevant for CARS are shown in Fig. \ref{im:exampleRealizationsBackground}.
%
%\begin{figure}
    %\centering
    %\includegraphics[width = %\textwidth]{pics/exampleErrorFunctions.png}
%    \caption{Example realizations for the error function $\varepsilon_{\rm{m}}(\bm\nu \mid \mu_e, \bm\theta)$ with $\mu_e = 1$, $\bm\theta = ( 0.08, 400)$ and starting and ending points $\widetilde{\bm\varepsilon}_{\rm ss} = ( 0.85, 1.02)^\intercal$.}
%    \label{im:exampleErrorRealizations}
%\end{figure}
%
\FloatBarrier

%%%%%%%%%%%%%%%%%%%%%%%%%%%%%%%%%%%%%%%%
\section{Raman and CARS spectrum models}
\label{sec:forwardModels}
In the preceding two sections we formulated mathematical procedures to sample synthetic spectrum and background realizations which are statistically similar to measurement data.
Below, we combine these two approaches for generating arbitrary amounts of statistically realistic spectral data which are ultimately used for training our Bayesian neural networks.
We present two forward models which are used to generate data for Raman measurements with an additive background and CARS measurements with a multiplicative background.

Raman spectra $y( \bm\nu )$ with an additive background $B( \bm\nu )$ are constructed using
\begin{equation}
    y( \bm\nu ) \sim r(\bm\nu \mid A, \bm\psi) + B( \bm\nu ),
    \label{eq:raman_model}
\end{equation}
where $r(\bm\nu \mid A, \bm\psi)$ is distributed according to the model defined in Eq.~\eqref{eq:lggp_statistical_spectrum_model}.
The background $B( \bm\nu )$ is sampled with Eq.~\eqref{eq:background_statistical_model}.

\begin{figure}
    \centering
    \includegraphics[width = 0.5\textwidth]{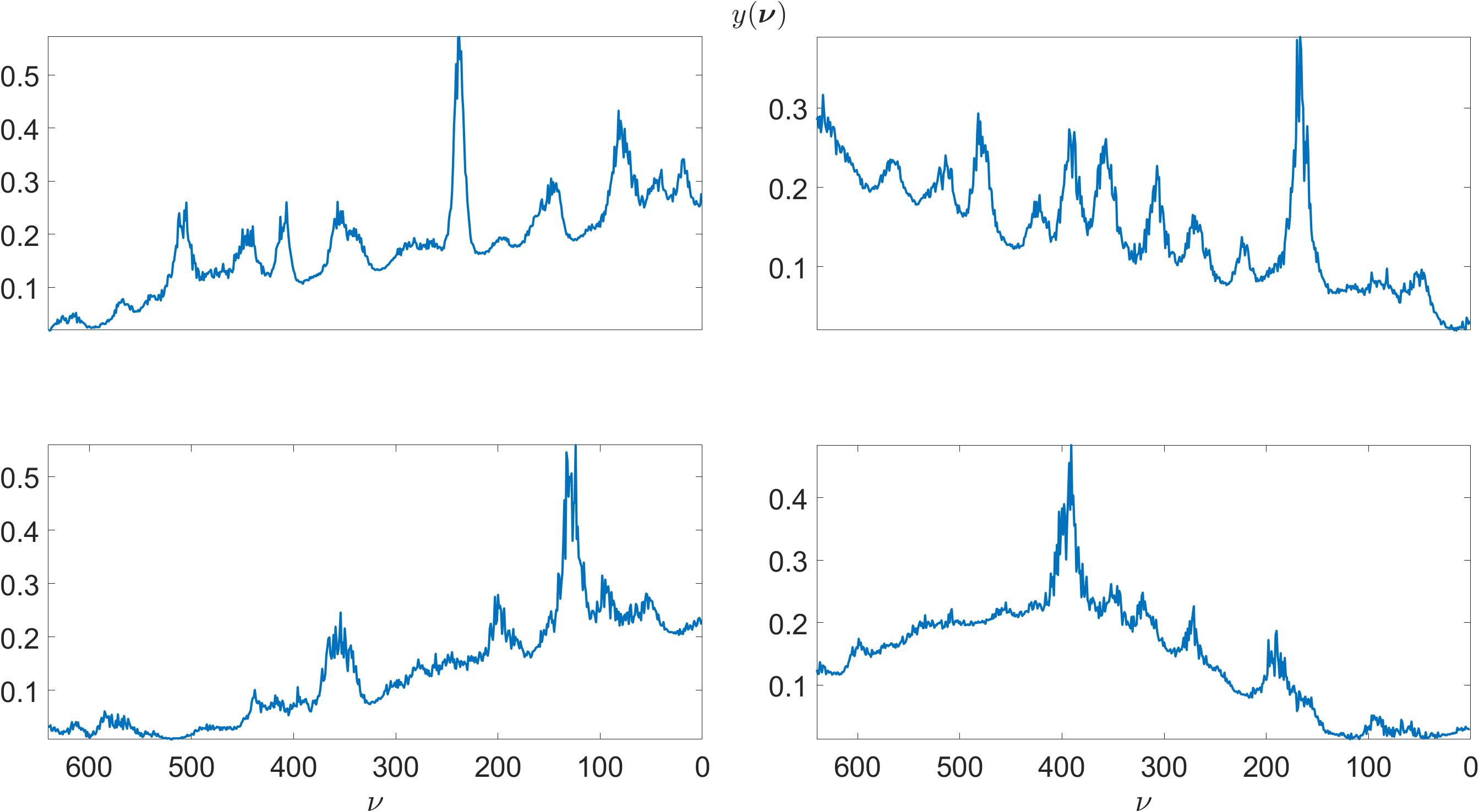}
    \caption{Example realizations for the Raman spectrum model defined in Eq.~\eqref{eq:raman_model}. The realizations correspond to the log-Gaussian gamma process realizations in Fig. \ref{im:exampleRealizationsLGGP}.}
    \label{im:exampleRealizationsRaman}
\end{figure}

\begin{figure}
    \centering
    \includegraphics[width = 0.5\textwidth]{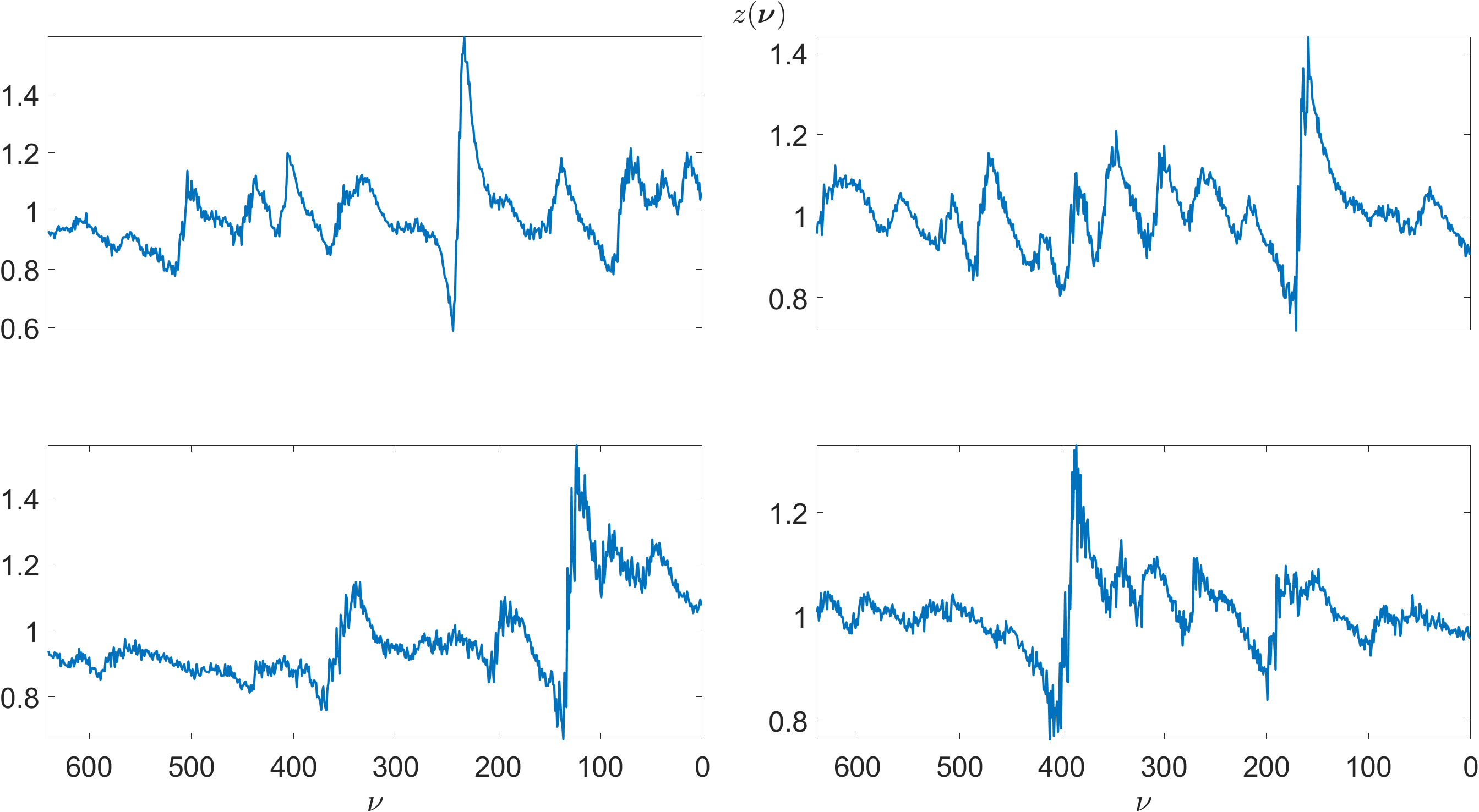}
    \caption{Example realizations for the CARS spectrum model defined in Eq.~\eqref{eq:cars_model}. The realizations correspond to the log-Gaussian gamma process realizations in Fig. \ref{im:exampleRealizationsLGGP}.}
    \label{im:exampleRealizationsCARS}
\end{figure}

CARS spectra $z( \bm\nu )$ are generated similarly to the additive Raman realizations.
The CARS model consists of a multiplicative background function $\varepsilon_{\rm{m}}(\bm\nu \mid \mu_e, \bm\theta_e)$ distorting a %corrected
CARS spectrum $S( \bm\nu \mid  B_{ \rm NR}, \bm\psi)$ given as  
\begin{equation}
    z( \bm\nu ) \sim \varepsilon_{\rm{m}}(\bm\nu \mid \mu_e, \bm\theta_e)S( \bm\nu \mid  B_{ \rm NR}, \bm\psi),
    \label{eq:cars_model}
\end{equation}
where the %corrected
CARS spectrum $S( \bm\nu \mid  B_{ \rm NR}, \bm\psi)$ can be given as
\begin{equation}
    S( \bm\nu \mid B_{ \rm NR}, \bm\psi) \sim \left\vert B_{ \rm NR} + \left( ir( \bm\nu \mid A, \bm\psi )-\mathcal{H}\left\{r( \bm\nu \mid A, \bm\psi) \right\}\right) \right\vert^2,
    \label{eq:cars_corrected_model}
\end{equation}
and $B_{ \rm NR} \sim \pi_0( B_{ \rm NR} )$ is a non-resonant background inherent to the CARS phenomenon distributed according to a prior distribution $\pi_0( B_{ \rm NR} )$ and $\mathcal{H}$ denotes the Hilbert transform.
The model for the CARS spectrum has been previously used for example in \cite{Kan:16, Harkonen:2020}.
We show example realizations for the Raman model in Fig. \ref{im:exampleRealizationsRaman} and the CARS model in Fig. \ref{im:exampleRealizationsCARS}.
We use the two models defined in Eqs.~\eqref{eq:raman_model} and \eqref{eq:cars_model} to generate two synthetic data sets which are used to train two separate Bayesian neural networks.
In the following section, we discuss the Bayesian neural network architecture.
\FloatBarrier

%%%%%%%%%%%%%%%%%%%%%%%%%%%%%%%%%%%%%%%%
\section{Bayesian neural network architecture}
\label{sec:networkArchitecture}
Our neural network architecture used in the experiments is based on the SpecNet architecture~\cite{Valensise:2020}.
The SpecNet architecture is composed of convolutional layers encoding the input, the measurement spectrum.
The encoded information is then decoded using fully-connected hidden layers, resulting in estimates for the underlying true Raman spectrum.
We present our changes to the SpecNet architecture below.

\begin{figure}
    \centering
    \includegraphics[width = 0.5\textwidth]{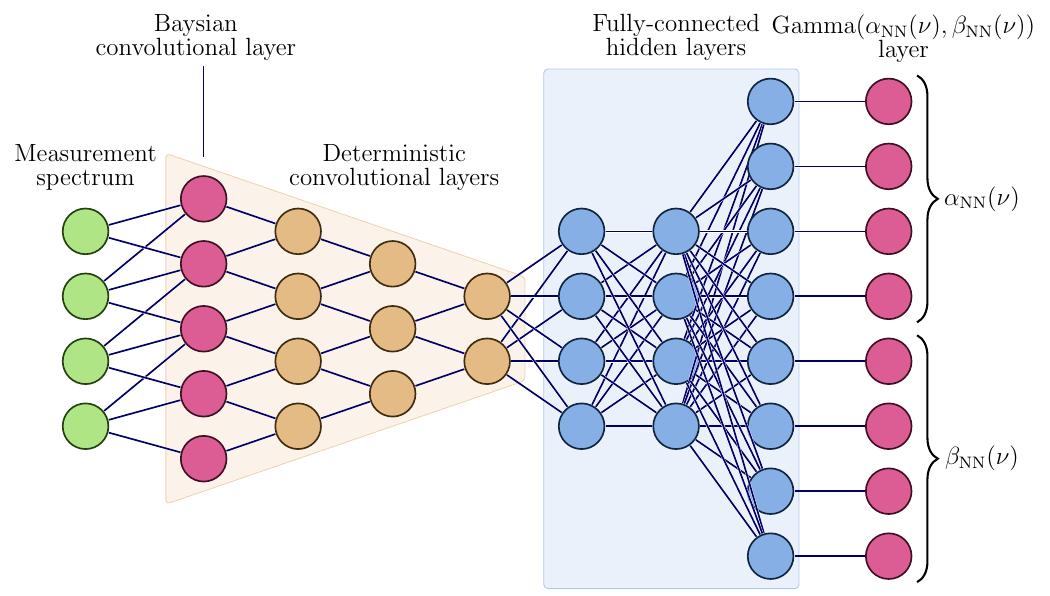}
    \caption{A Bayesian neural network architecture for correcting spectral measurements. The first convolutional layer is modeled as a stochastic layer. A new representation of the measurement spectrum is produced via the convolutional layers and then decoded with the fully-connected hidden layers. The output layer is modeled as a gamma process $\rm{Gamma}( \alpha_{\rm NN}(\nu), \beta_{\rm NN}(\nu))$, parameters of which are the outputs of the fully-connected hidden layers.}
    \label{im:gammaSpecnet}
\end{figure}

To achieve a partially Bayesian neural network~\cite{Sharma:2023}, we use a Bayesian layer for the first convolutional layer.
Additionally, we augment the architecture with a probabilistic output layer.
This transforms the neural network estimate into estimates of a stochastic process instead of directly estimating the Raman spectrum.
We use a gamma distribution as our output layer, following our formulation of Raman spectra as a log-Gaussian gamma processes.
We also found that $L_1$ or $L_2$ regularization was not necessary for the deterministic parts of the network and therefore only employ Dropout\cite{Srivastava:2014} regularization with the last dense layer of the network.
This in agreement with the documented robustness of Bayesian neural networks with respect to overfitting~\cite{Magris:2023}.
The above results in the following partial posterior probability distribution, or cost function, used for training the neural network
\begin{equation}
    \pi( \Psi_{ \rm D }, \Psi_{ \rm S } \mid R ) \propto \mathcal{L}( R \mid \Psi_{ \rm D }, \Psi_{ \rm S } ) \pi_0( \Psi_{ \rm S } ),
    \label{eq:cnnPosterior}
\end{equation}
where $R \in \mathbb{R}^{ I \times J }$ is a data matrix of $I$ synthetic spectra of length $J$ generated using either the Raman or CARS forward models in Eqs.~\eqref{eq:raman_model} and \eqref{eq:cars_model} and $\mathcal{L}( R \mid \Psi_{ \rm D }, \Psi_{ \rm S } )$ denotes the likelihood of the neural network estimate and $\pi_0( \Psi_{ \rm S } )$ is the prior distribution for the stochastic parameters of the network.
As our outputs are modeled as gamma-distributed random variables, the likelihood $\mathcal{L}( R \mid \Psi_{ \rm D }, \Psi_{ \rm S } )$ is given as
\begin{equation}
    \mathcal{L}( R \mid \Psi_{ \rm D }, \Psi_{ \rm S } ) = \prod\limits_{i = 1}^I \prod\limits_{ j = 1}^J \frac{ R_{i,j}^{\alpha_{\mathrm{NN},j} - 1}\exp( -R_{i,j}/\beta_{\mathrm{NN},j} ) }{ \Gamma(\alpha_{\mathrm{NN},j}  ) \beta_{\mathrm{NN},j}^{\alpha_{\mathrm{NN},j}} },
\end{equation}
where $R_{i,j}$ denotes the $j$th data point of the $j$th spectrum, $\alpha_{\mathrm{NN},j}$ and $\beta_{\mathrm{NN},j}$ are neural network outputs for the gamma distribution parameters.
For the prior distribution $\pi_0( \Psi_{ \rm S } )$, we use an independent normal distribution $\mathcal{N}( 0, 1)$ for all the weights and biases of the first layer, $\pi_0( \Psi_{ \rm S } ) \propto \prod_{p = 1}^P \mathcal{N}( \Psi_{ \mathrm{S}, p}; 0, 1)$ where $P$ is the total number of parameters in the first layer and $\mathcal{N}( \Psi_{ \mathrm{S}, p}; 0, 1)$ denotes the evaluation of the probability density at the parameter value $\Psi_{ \mathrm{S}, p}$.
This particular type of Bayesian neural network is also known as a deep kernel \cite{Wilson:2016} or a manifold Gaussian processes \cite{Calandra:2016}.
We illustrate the neural network architecture in Fig. \ref{im:gammaSpecnet}.
%
%
%\FloatBarrier

In the log-Gaussian gamma process section, we estimate parameters of a doubly-stochastic process via MCMC.
The Bayesian neural network architecture proposed here can be seen as an estimate of a \textit{triply}-stochastic process where the neural network outputs are two stochastic process realizations $\alpha_{\mathrm{NN}}(\bm\nu)$ and $\beta_{\mathrm{NN}}(\bm\nu)$, an extension to the analytical log-Gaussian gamma process in Section~\ref{sec:lggp_model} where the log-Gaussian parameterization of the scale process $\beta_{\mathrm{NN}}(\bm\nu)$ is used for mathematical convenience due to the closed form of the probability density in Eq.~\eqref{eq:beta_gp_loglikelihood}.

The uncertainty quantification of the Bayesian neural network is achieved in two stages, the Bayesian convolutional layer and gamma distributed output layer.
Numerical samples are generated for the parameters of the Bayesian convolutional layer during the training process.
These numerical samples are propagated through the deterministic layers of the network and ultimately into the output layer.
The output layer, modelled as a gamma process, outputs a gamma distribution for each prediction point which ultimately controls the uncertainty of the spectrum estimate.

\section{Computational details and prior distributions}
\label{sec:priors}
\begin{figure}
    \centering
    \includegraphics[width = 0.5\textwidth]{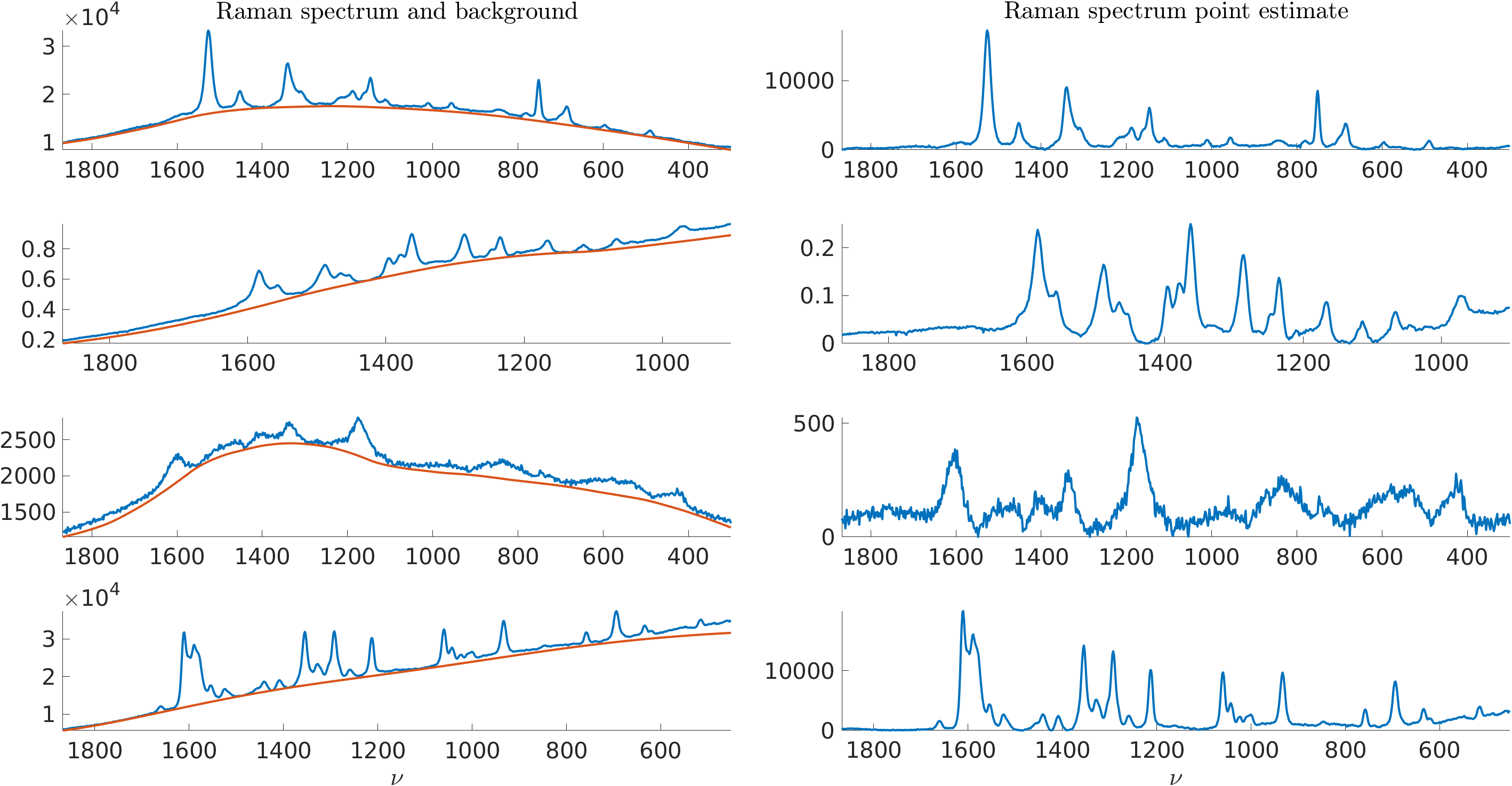}
    \caption{On the left, experimental Raman spectra of phthalocyanine blue, naphthol red, aniline black, and red 264 pigments in blue and point estimates for their respective additive backgrounds $B(\bm\nu)$. On the right, point estimates for the underlying Raman spectra corresponding to the Raman measurements on their left. The 4 cases are used to estimate the LGGP and GP parameters defined in the posterior distributions in Eqs.~\eqref{eq:lggp_statistical_spectrum_model} and \eqref{eq:background_statistical_model}.}
    \label{im:ramanData}
\end{figure}

\begin{figure}
    \centering
    \includegraphics[width = 0.5\textwidth]{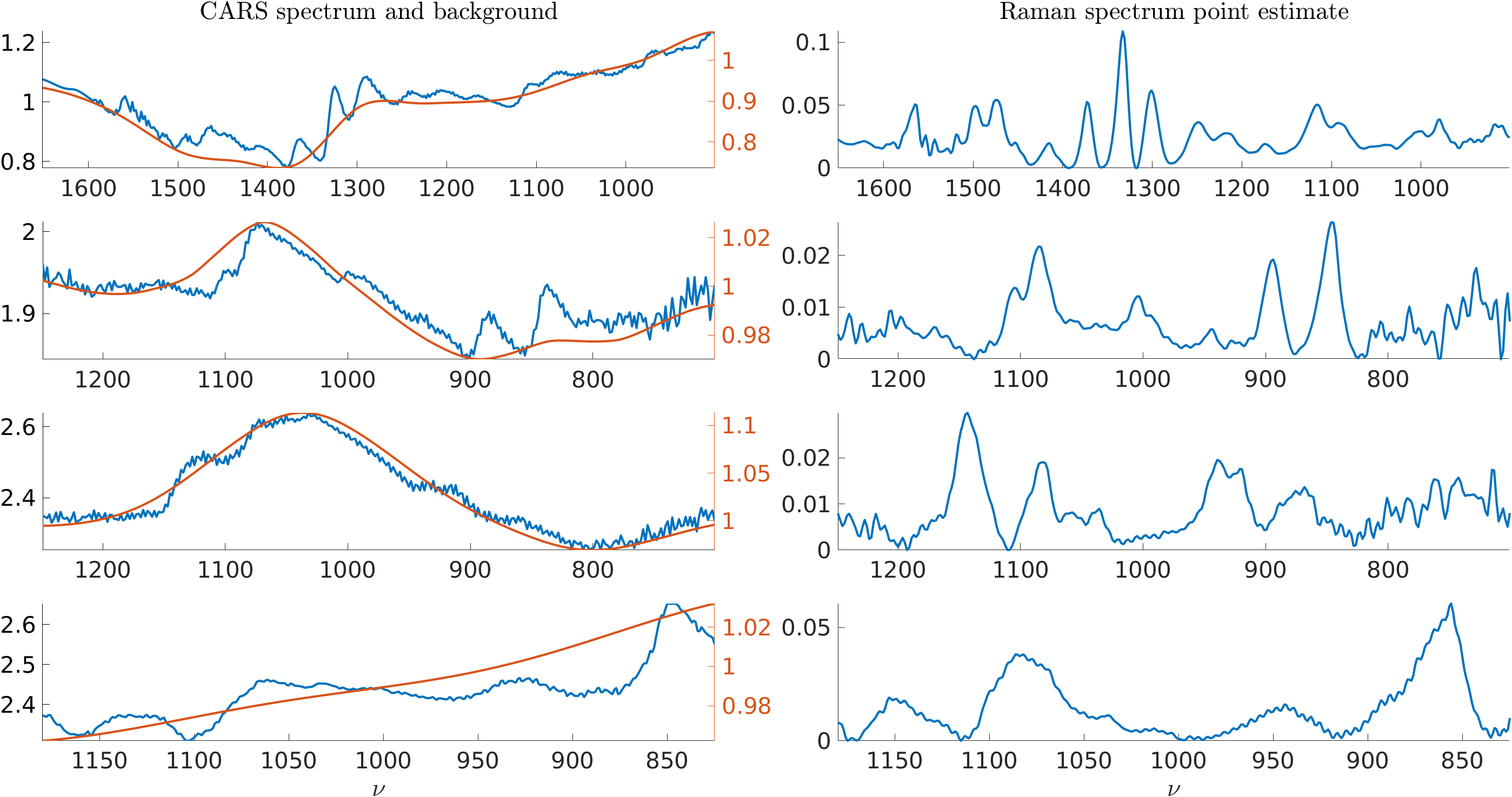}
    \caption{On the left, experimental CARS spectra of adenosine phosphate, fructose, glucose, and sucrose in blue and point estimates for their respective multiplicative backgrounds $\varepsilon_{\rm{m}}(\bm\nu)$. On the right, point estimates for the underlying Raman spectra corresponding to the CARS measurements on their left. The 4 cases are used to estimate the LGGP and GP parameters defined in the posterior distributions in Eqs.~\eqref{eq:lggp_statistical_spectrum_model} and \eqref{eq:background_statistical_model}.}
    \label{im:carsData}
\end{figure}

We use 4 experimental Raman spectra and 4 CARS spectra to generate the synthetic training data sets.
We use a wavelet-based approach\cite{Harkonen:2023} to obtain point estimates for the underlying Raman spectra in all 8 cases.
Additionally, the method provides point estimates for the additive and multiplicative background signal which we use to estimate the parameters of the background GP model defined in Eq.~\eqref{eq:background_statistical_model}.
We show the obtained Raman data point estimates for the Raman spectra and additive backgrounds in Fig.~\ref{im:ramanData} and
CARS point estimates for the Raman spectra and multiplicative backgrounds in Fig.~\ref{im:carsData}.
The four cases of measurement data are used to train their respective Bayesian neural network architectures.
It should be noted that for cases with significantly different Raman spectral signatures, such as where the spectra consists of either significantly sharper or wider line shapes, the training should be done using experimental data which contain such features.

We run the DRAM algorithm with 5 proposal steps and with a length of 100\,000 samples for both the log-Gaussian gamma process parameters and the GP parameters.
We use a burn-in of 50\,000 samples.
The prior distributions for the log-Gaussian gamma process likelihood and the GP background likelihood are documented in Table \ref{tb:priors}.
We use TensorFlow and TensorFlow Probability together with Keras to implement the neural network architecture \cite{Tensorflow:2015, Dillon:2017, Chollet:2015}.
We use the Adam optimizer for estimating the network parameters $  \Psi_{ \rm D }$ and $\Psi_{ \rm S } $.

The aforementioned MCMC sampling runs were done on an AMD Ryzen 3950X processor.
Wall times for the MCMC sampling ranged from a couple of minutes to approximately one hour, depending on the number of data points in the spectra.
The MCMC samplers can be run embarrassingly parallel for each measurement spectrum.
Training the Bayesian neural network took approximately two hours on an NVIDIA 1070 graphics card for sets of 500 000 spectra.
Given the small computational cost of the MCMC sampling and training the neural network, which are offline costs, the main computational limitation comes from loading data during inference.

\begin{table}
\caption{Prior distributions for the log-Gaussian gamma and GP parameters.}
\centering
\begin{tabular}{c c c c}
\toprule
Parameter & Distribution & Parameter & Distribution \\
\midrule
$\pi_0( \alpha )$ & $\mathcal{U}( 0, \infty)$ & & \\
$\pi_0( c_\beta )$ & $\mathcal{U}( 0, \infty)$ & $\pi_0( \sigma_G )$ & $\mathcal{U}( 1, \infty)$ \\
$\pi_0( \mu_\beta )$ & $\mathcal{U}( 0, \infty)$ & $\pi_0( \mu_e )$ & $\mathcal{U}( 0, \infty)$ \\
$\pi_0( \sigma_{\beta,f} )$ & $\mathcal{U}( 0, \infty )$ & $\pi_0( \sigma_{e,f} )$ & $\mathcal{U}( 0, \infty)$ \\
$\pi_0( l_\beta )$ & $\mathcal{U}( 0, \infty)$ & $\pi_0( l_e )$ & $\mathcal{U}( 0, \infty)$ \\
$\pi_0( \sigma_\beta )$ & $\mathcal{U}( 0, \infty)$ & $\pi_0( \sigma_e )$ & $\mathcal{U}( 0, \infty)$ \\
$\pi_0( e_{\rm start} )$ & $\mathcal{U}( 0.90, 1.10)$ & $\pi_0( e_{\rm stop} )$ & $\mathcal{U}( 0.90, 1.10)$ \\
\bottomrule
\end{tabular}
\label{tb:priors}
\end{table}
%
%
%%%%%%%%%%%%%%%%%%%%%%%%%%%%%%%%%%%%%%%%
\section{Results}
\label{sec:results}
\begin{figure}
    \centering
    \includegraphics[width = 0.5\textwidth]{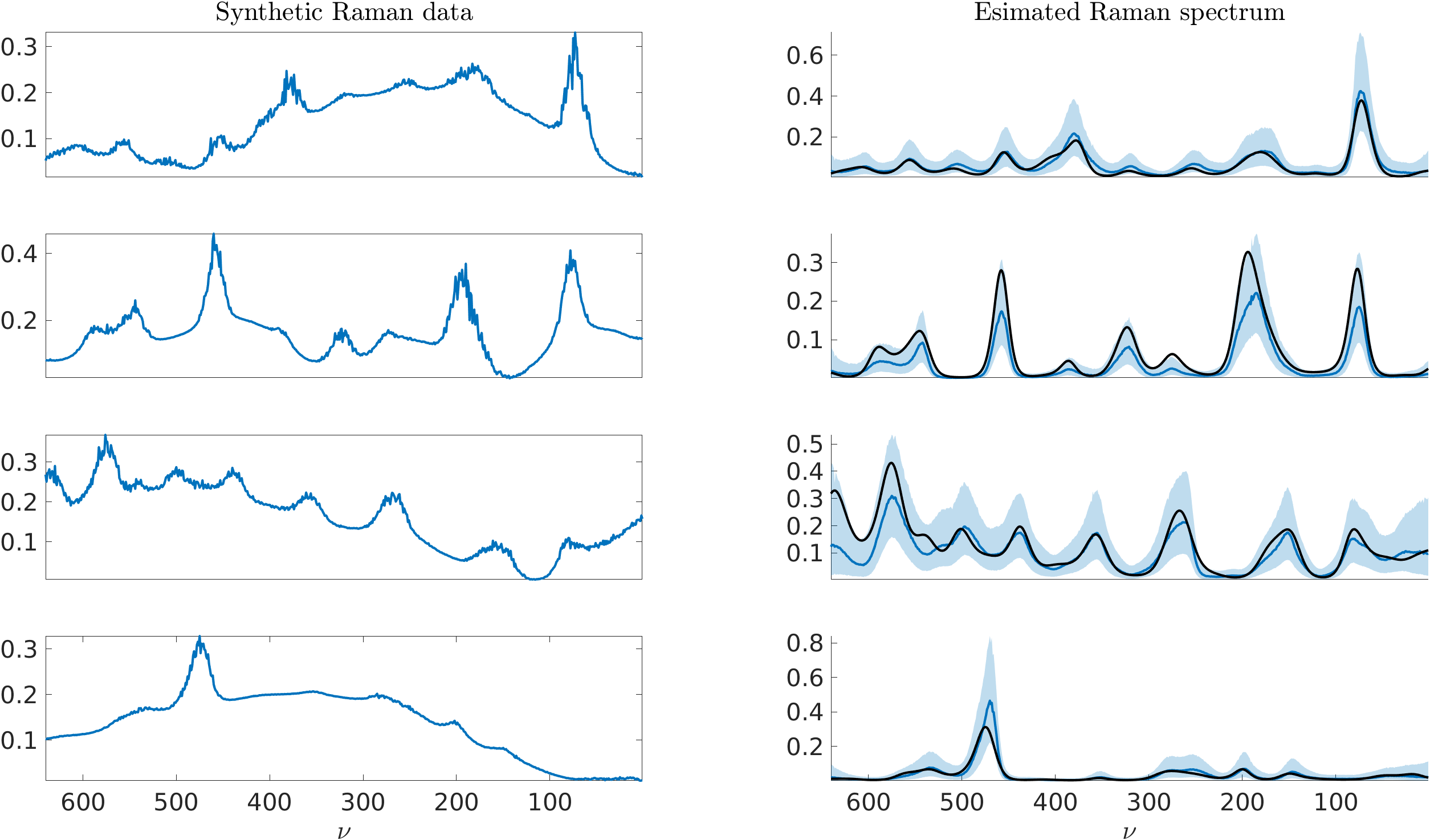}
    \caption{On the left, synthetic Raman test spectra generated using Eq.~\eqref{eq:raman_model}. On the right, corresponding Raman spectrum estimates with the median estimate shown in solid blue and the 90\% confidence interval in shaded blue. The solid black line shows the ground truth spectrum.}
    \label{im:syntheticResultsRaman}
\end{figure}

\begin{figure}
    \centering
    \includegraphics[width = 0.5\textwidth]{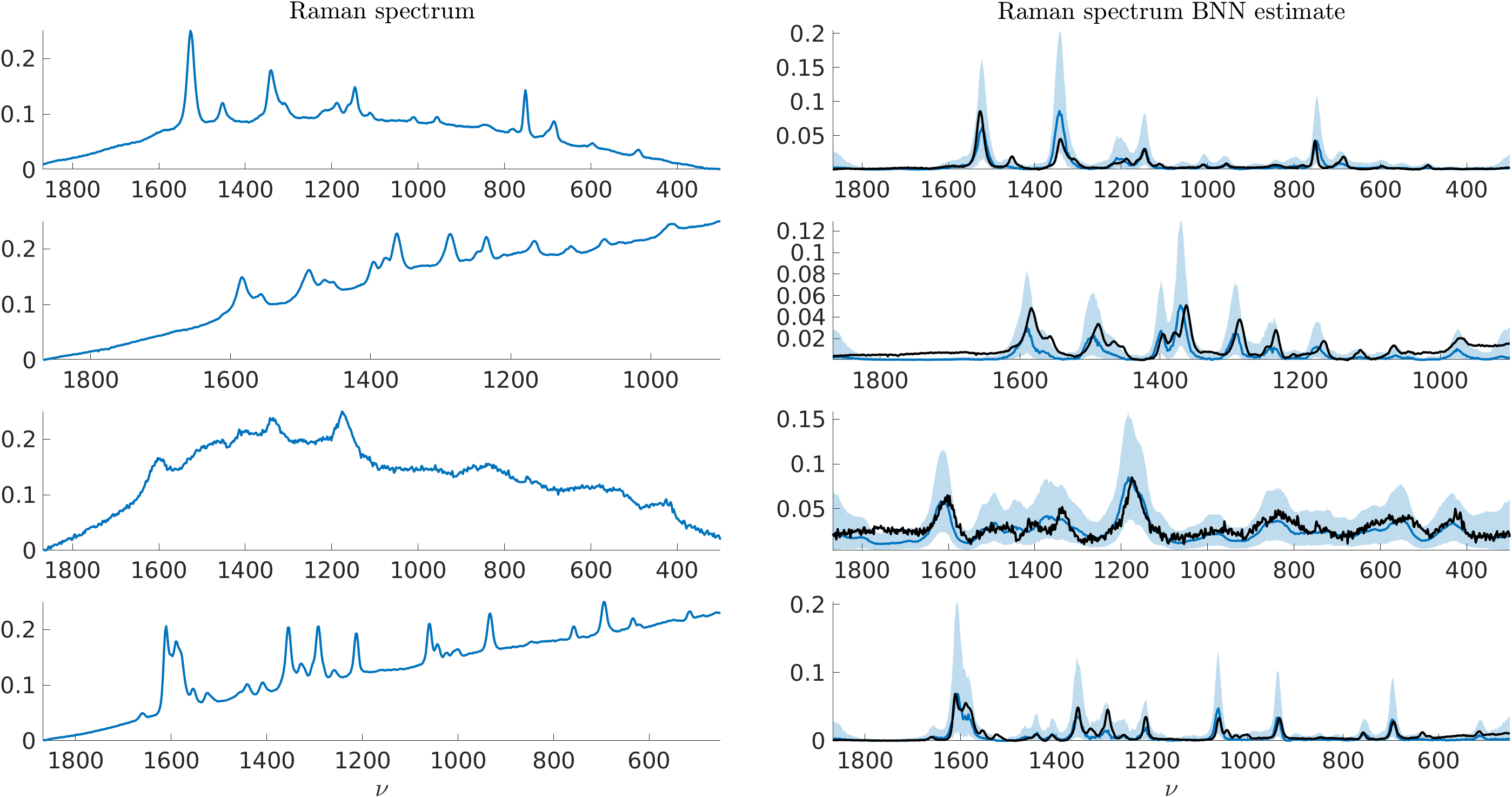}
    \caption{On the left, experimental Raman spectra of phthalocyanine blue, aniline black, naphthol red, and red 264 pigments. On the right, corresponding Raman spectrum estimates with the median estimate shown in solid blue and the 90\% confidence interval in shaded blue. The solid black line shows the ground truth spectrum.}
    \label{im:experimentalResultsRaman}
\end{figure}

We apply the two Bayesian neural networks to 4 synthetic Raman spectra and 4 synthetic CARS spectra generated using Eqs.~\eqref{eq:raman_model} and \eqref{eq:cars_model}, respectively.
The synthetic spectra were not part of the training data sets.
The synthetic data and the results for Raman spectra with an additive background are shown in Fig. \ref{im:syntheticResultsRaman} and results for experimental Raman spectra of phthalocyanine blue, naphthol red, aniline black, and red 264 pigments are presented in Fig. \ref{im:experimentalResultsRaman}.
The experimental details of the CARS samples have been described in detail elsewhere, see for example \cite{Harkonen:2020}.
The Raman spectra are from an online database of Raman spectra of pigments used in
modern and contemporary art (The standard Pigments Checker v.5) \cite{ramanDatabase}.

\begin{figure}
    \centering
    \includegraphics[width = 0.5\textwidth]{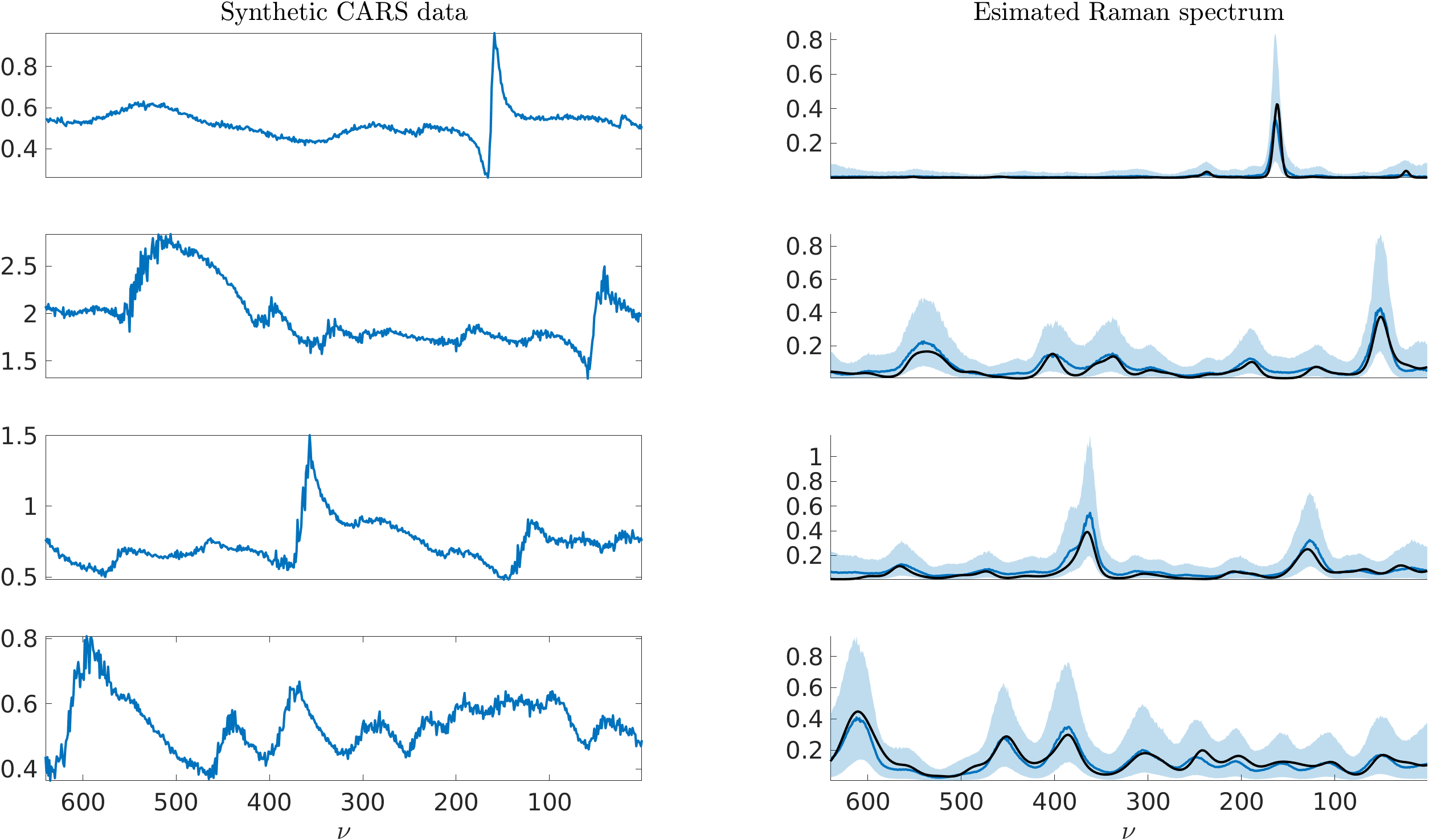}
    \caption{On the left, synthetic CARS test spectra generated using Eq.~\eqref{eq:cars_model}. On the right, corresponding Raman spectrum estimates with the median estimate shown in solid blue and the 90\% confidence interval in shaded blue. The solid black line shows the ground truth spectrum.}
    \label{im:syntheticResultsCARS}
\end{figure}

\begin{figure}
    \centering
    \includegraphics[width = 0.5\textwidth]{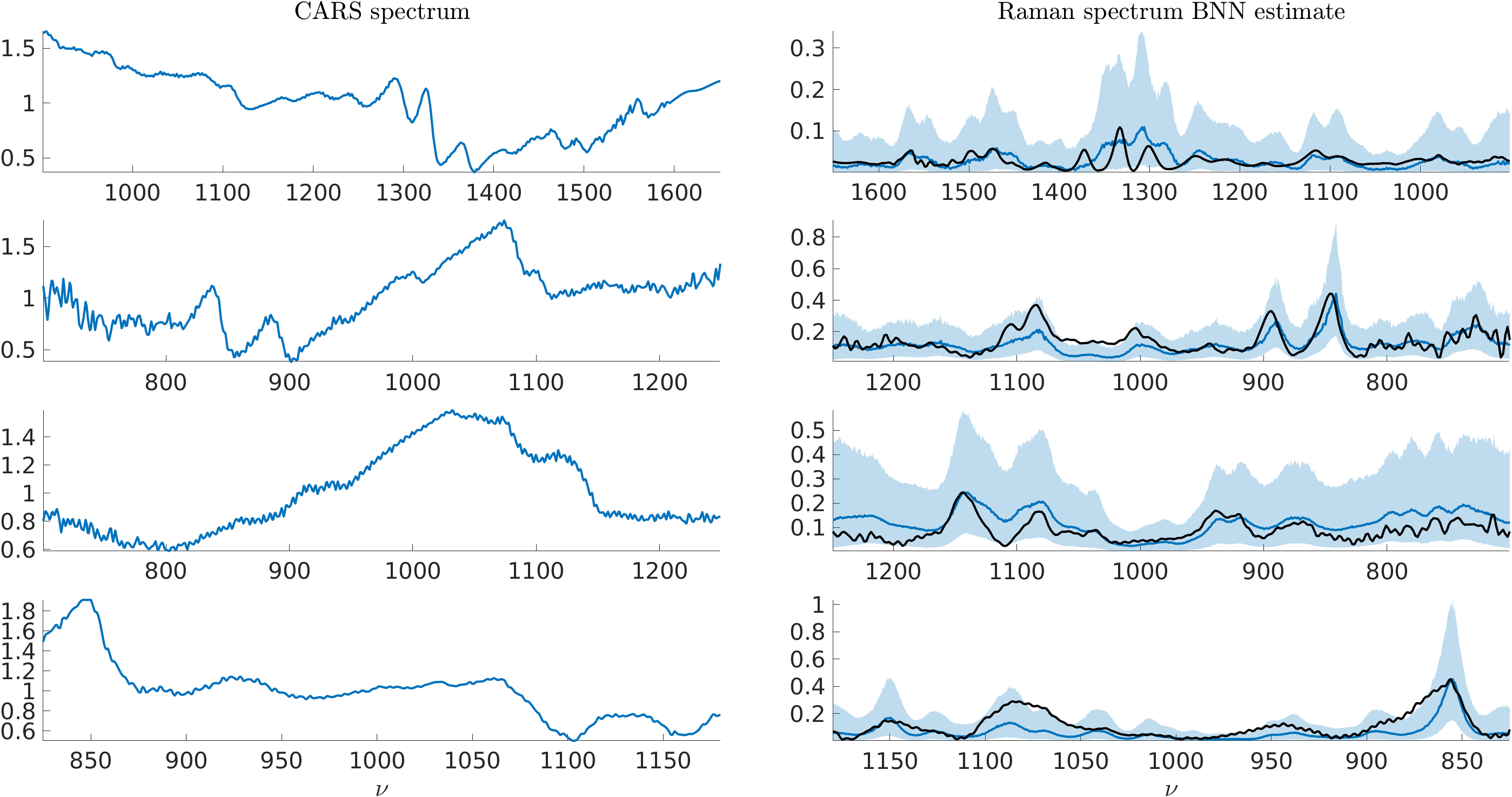}
    \caption{On the left, experimental CARS spectra of adenosine phosphate, fructose, glucose, and sucrose. On the right, corresponding Raman spectrum estimates with the median estimate shown in solid blue and the 90\% confidence interval in shaded blue. The solid black line shows the ground truth spectrum.}
    \label{im:experimentalResultsCARS}
\end{figure}

Results for synthetic CARS spectra are shown in Fig. \ref{im:syntheticResultsCARS} and results for experimental CARS spectra of adenosine phosphate, fructose, glucose, and sucrose are presented in Fig. \ref{im:experimentalResultsCARS}.
The spectra themselves were not part of the training data set.
The results show the median estimate of the Raman spectrum obtained from the trained Bayesian neural network along with the 90\% confidence intervals of the Raman spectrum estimate.
We overlay the Raman spectrum estimate with a scaled versions of the point estimates in Fig. \ref{im:carsData}.
The point estimates are scaled such that the minima and maxima of the point estimate are equal to the minima and maxima of the median estimate of the Raman spectrum.
The results coincide with the overall shape of the point estimates, supporting the validity of the data generation approach and the Bayesian neural network design.

As additional validation for the synthetic spectrum generation approach, we compare the cost function values, see Eq.~\eqref{eq:lggp_posterior}, of the fully synthetic training spectra to the cost function values obtained for partially experimental spectra.
We generate the partially experimental spectra by taking the point estimates of the experimental CARS spectra in Figure \ref{im:carsData} and generating sets of CARS spectra with the forward model defined in Eq.~\eqref{eq:cars_model} combined with the GP realizations.
The results are in agreement which implies that our log-Gaussian gamma process is capable of generating valid Raman spectra for training neural networks.
This approach is similar to approaches used for approximate Bayesian computation, see for example \cite{Jiang:2017, Grazian:2020}.
The resulting log cost function distribution of the synthetic spectra and the partially experimental spectra are provided in the Electronic Supplementary Information.
%
%%%%%%%%%%%%%%%%%%%%%%%%%%%%%%%%%%%%%%%%
\section{Conclusions}
\label{sec:conclusions}
We propose a novel approach utilizing log-Gaussian gamma processes and Gaussian processes to generate synthetic spectra and additive or multiplicative backgrounds that are statistically similar to experimental measurements, even when using a limited number of experimental spectra.
The parameters of these stochastic processes are learned through Markov chain Monte Carlo methods, enabling the generation of extensive training data for neural networks by sampling from Bayesian posterior distributions of the parameters.
%
%We proposed using log-Gaussian gamma processes and GPs as a way to generate synthetic spectra and additive or multiplicative backgrounds which are statistically similar to experimental measurements.
%
%The method can be applied even with a small number of experimental spectra.
%
%We learned the parameters of the aforementioned stochastic processes using Markov chain Monte Carlo methods, yielding samples from Bayesian posterior distributions.
%
%Sampling the posterior distributions can be used to generate arbitrary amounts of training data for purposes of training neural networks.
%

This data generation method is applied to train two Bayesian neural networks, specifically designed for correcting spectral measurements.
One network is tailored for Raman spectra with additive backgrounds, while the other is optimized for coherent anti-Stokes Raman scattering (CARS) spectra with multiplicative backgrounds.
Bayesian neural networks expand upon prior research involving neural networks for spectral corrections, offering not only point estimates but also the critical capability of uncertainty quantification.

%We employed the proposed data generation approach to learn the parameters of two Bayesian neural networks for correcting spectral measurements of Raman spectra with additive backgrounds and coherent anti-Stokes Raman scattering spectra with multiplicative backgrounds.
%
%The use of Bayesian neural networks extends pre-existing literature on using neural networks for spectral corrections by allowing uncertainty quantification along with point estimates obtainable with non-Bayesian neural networks.
%

Our approach is validated using synthetic test data generated from the stochastic processes and experimental Raman spectra of phthalocyanine blue, aniline black, naphthol red, and red 264 pigments, along with experimental CARS spectra of adenosine phosphate, fructose, glucose, and sucrose.
The results demonstrate excellent agreement with deterministically obtained point estimates of the Raman spectra, while simultaneously providing valuable uncertainty estimates for the Raman spectrum estimates.

As a future avenue of research, our log-Gaussian gamma process formulation could be extended to a mixture of log-Gaussian gamma processes, similarly to mixtures of Gaussian processes \cite{tresp2001mixtures, Rasmussen:2002, Harkonen:2022:MoE}.
Such an extension would allow modelling of nonstationarity and heteroscedasticity, meaning different signal or noise behaviour at different parts of measurement spectrum.
A more straight-forward modification, if necessary, would be to include an additional noise term consisting of, for example, Gaussian white noise process to the log-Gaussian gamma process formulation.

In addition, log-Gaussian gamma processes might be used to model other inherently positive measurements such as reflectance, absorbance, fluorescence, or transmittance spectra \cite{Wittenberghe:2014, Lazaro-Gredilla:2014, Ghosh:2021} and other non-spectroscopic data sets like pollutant or protein concentrations or masses of stellar objects,  for which Gaussian processes have been used \cite{Lawrence:2006,  Bu:2022, Harkonen:2023:Plume}
Modifications to the log-Gaussian gamma process and the Bayesian neural network should be minimal due to the uninformative priors and general purpose structure of them.
The generation of synthetic data sets for other types of problems requires an appropriate forward model.
If the forward model has non-local behaviour, it might warrant architectural changes to the Bayesian neural network such as augmenting the first layer with a dense layer for modelling the non-local dependencies in the data.

A comparison study similar to \cite{Junjuri:2023} would be an interesting avenue of future research.
However, a direct comparison between non-Bayesian and Bayesian neural networks is not straight-forward due to the missing uncertainty quantification of the non-Bayesian neural network estimates.
One could possibly use the Dropout regularization as an approximate Bayesian approach\cite{Gal:2016}.
This could be combined with simulation-based calibration\cite{Talts:2020} to provide a appropriate one-to-one comparison between the non-Bayesian and Bayesian neural networks.

%We validated the approach with synthetic test data generated from the stochastic processes along with experimental Raman spectra of phthalocyanine blue, aniline black, naphthol red, and red 264 pigments and experimental CARS spectra of adenosine phosphate, fructose, glucose, and sucrose.
%
%The results were in agreement with deterministically obtained point estimates of the Raman spectra while simultaneously having uncertainty estimates for the Raman spectrum estimates.
%

%%%%%%%%%%%%%%%%%%%%%%%%%%%%%%%%%%%%%%%%
\section*{Author Contributions}
T. H.: conceptualization, formal analysis, investigation, methodology, software, visualization. E. M. V. and L. L: resources. M. T. M.: investigation, methodology. L. R.: funding acquisition, project administration, supervision. All authors: writing – original draft, writing – review \& editing.
%%%%%%%%%%%%%%%%%%%%%%%%%%%%%%%%%%%%%%%%

%%%%%%%%%%%%%%%%%%%%%%%%%%%%%%%%%%%%%%%%
\section*{Conflicts of interest}
There are no conflicts to declare.

%%%%%%%%%%%%%%%%%%%%%%%%%%%%%%%%%%%%%%%%
\section*{Acknowledgements}
The authors were supported by Research Council of Finland (grant number 353095). T. H. thanks Ilmari Vahteristo for the fruitful discussions.
%
%%%END OF MAIN TEXT%%%

%The \balance command can be used to balance the columns on the final page if desired. It should be placed anywhere within the first column of the last page.

\balance

%If notes are included in your references you can change the title from 'References' to 'Notes and references' using the following command:
%\renewcommand\refname{Notes and references}

%%%REFERENCES%%%
\bibliography{rsc} %You need to replace "rsc" on this line with the name of your .bib file

\providecommand*{\mcitethebibliography}{\thebibliography}
\csname @ifundefined\endcsname{endmcitethebibliography}
{\let\endmcitethebibliography\endthebibliography}{}
\begin{mcitethebibliography}{72}
\providecommand*{\natexlab}[1]{#1}
\providecommand*{\mciteSetBstSublistMode}[1]{}
\providecommand*{\mciteSetBstMaxWidthForm}[2]{}
\providecommand*{\mciteBstWouldAddEndPuncttrue}
  {\def\EndOfBibitem{\unskip.}}
\providecommand*{\mciteBstWouldAddEndPunctfalse}
  {\let\EndOfBibitem\relax}
\providecommand*{\mciteSetBstMidEndSepPunct}[3]{}
\providecommand*{\mciteSetBstSublistLabelBeginEnd}[3]{}
\providecommand*{\EndOfBibitem}{}
\mciteSetBstSublistMode{f}
\mciteSetBstMaxWidthForm{subitem}
{(\emph{\alph{mcitesubitemcount}})}
\mciteSetBstSublistLabelBeginEnd{\mcitemaxwidthsubitemform\space}
{\relax}{\relax}

\bibitem[Mulvaney and Keating(2000)]{Mulvaney:2000}
S.~P. Mulvaney and C.~D. Keating, \emph{Analytical Chemistry}, 2000, \textbf{72}, 145--158\relax
\mciteBstWouldAddEndPuncttrue
\mciteSetBstMidEndSepPunct{\mcitedefaultmidpunct}
{\mcitedefaultendpunct}{\mcitedefaultseppunct}\relax
\EndOfBibitem
\bibitem[Krafft \emph{et~al.}(2012)Krafft, Dietzek, Popp, and Schmitt]{Krafft:2012}
C.~Krafft, B.~Dietzek, J.~Popp and M.~Schmitt, \emph{Journal of Biomedical Optics}, 2012, \textbf{17}, 040801\relax
\mciteBstWouldAddEndPuncttrue
\mciteSetBstMidEndSepPunct{\mcitedefaultmidpunct}
{\mcitedefaultendpunct}{\mcitedefaultseppunct}\relax
\EndOfBibitem
\bibitem[Day \emph{et~al.}(2011)Day, Domke, Rago, Kano, Hamaguchi, Vartiainen, and Bonn]{Day:2011}
J.~P.~R. Day, K.~F. Domke, G.~Rago, H.~Kano, H.-o. Hamaguchi, E.~M. Vartiainen and M.~Bonn, \emph{The Journal of Physical Chemistry B}, 2011, \textbf{115}, 7713--7725\relax
\mciteBstWouldAddEndPuncttrue
\mciteSetBstMidEndSepPunct{\mcitedefaultmidpunct}
{\mcitedefaultendpunct}{\mcitedefaultseppunct}\relax
\EndOfBibitem
\bibitem[Boelens \emph{et~al.}(2005)Boelens, Eilers, and Hankemeier]{Boelens:2005}
H.~F.~M. Boelens, P.~H.~C. Eilers and T.~Hankemeier, \emph{Analytical Chemistry}, 2005, \textbf{77}, 7998--8007\relax
\mciteBstWouldAddEndPuncttrue
\mciteSetBstMidEndSepPunct{\mcitedefaultmidpunct}
{\mcitedefaultendpunct}{\mcitedefaultseppunct}\relax
\EndOfBibitem
\bibitem[He \emph{et~al.}(2014)He, Zhang, Liu, Huang, He, Xie, Wu, and Du]{He:2014}
S.~He, W.~Zhang, L.~Liu, Y.~Huang, J.~He, W.~Xie, P.~Wu and C.~Du, \emph{Anal. Methods}, 2014, \textbf{6}, 4402--4407\relax
\mciteBstWouldAddEndPuncttrue
\mciteSetBstMidEndSepPunct{\mcitedefaultmidpunct}
{\mcitedefaultendpunct}{\mcitedefaultseppunct}\relax
\EndOfBibitem
\bibitem[Gan \emph{et~al.}(2006)Gan, Ruan, and Mo]{Gan:2006}
F.~Gan, G.~Ruan and J.~Mo, \emph{Chemometrics and Intelligent Laboratory Systems}, 2006, \textbf{82}, 59--65\relax
\mciteBstWouldAddEndPuncttrue
\mciteSetBstMidEndSepPunct{\mcitedefaultmidpunct}
{\mcitedefaultendpunct}{\mcitedefaultseppunct}\relax
\EndOfBibitem
\bibitem[Galloway \emph{et~al.}(2009)Galloway, Ru, and Etchegoin]{Galloway:09}
C.~M. Galloway, E.~C.~L. Ru and P.~G. Etchegoin, \emph{Appl. Spectrosc.}, 2009, \textbf{63}, 1370--1376\relax
\mciteBstWouldAddEndPuncttrue
\mciteSetBstMidEndSepPunct{\mcitedefaultmidpunct}
{\mcitedefaultendpunct}{\mcitedefaultseppunct}\relax
\EndOfBibitem
\bibitem[Vartiainen \emph{et~al.}(2006)Vartiainen, Rinia, M\"{u}ller, and Bonn]{Vartiainen:06}
E.~M. Vartiainen, H.~A. Rinia, M.~M\"{u}ller and M.~Bonn, \emph{Opt. Express}, 2006, \textbf{14}, 3622--3630\relax
\mciteBstWouldAddEndPuncttrue
\mciteSetBstMidEndSepPunct{\mcitedefaultmidpunct}
{\mcitedefaultendpunct}{\mcitedefaultseppunct}\relax
\EndOfBibitem
\bibitem[Kan \emph{et~al.}(2016)Kan, Lensu, Hehl, Volkmer, and Vartiainen]{Kan:16}
Y.~Kan, L.~Lensu, G.~Hehl, A.~Volkmer and E.~M. Vartiainen, \emph{Opt. Express}, 2016, \textbf{24}, 11905--11916\relax
\mciteBstWouldAddEndPuncttrue
\mciteSetBstMidEndSepPunct{\mcitedefaultmidpunct}
{\mcitedefaultendpunct}{\mcitedefaultseppunct}\relax
\EndOfBibitem
\bibitem[Chi \emph{et~al.}(2019)Chi, Han, Xu, Wang, Shu, Zhou, and Wu]{Chi:2019}
M.~Chi, X.~Han, Y.~Xu, Y.~Wang, F.~Shu, W.~Zhou and Y.~Wu, \emph{Applied Spectroscopy}, 2019, \textbf{73}, 78--87\relax
\mciteBstWouldAddEndPuncttrue
\mciteSetBstMidEndSepPunct{\mcitedefaultmidpunct}
{\mcitedefaultendpunct}{\mcitedefaultseppunct}\relax
\EndOfBibitem
\bibitem[H\"{a}rk\"{o}nen and Vartiainen(2023)]{Harkonen:2023}
T.~H\"{a}rk\"{o}nen and E.~Vartiainen, \emph{Opt. Continuum}, 2023, \textbf{2}, 1068--1076\relax
\mciteBstWouldAddEndPuncttrue
\mciteSetBstMidEndSepPunct{\mcitedefaultmidpunct}
{\mcitedefaultendpunct}{\mcitedefaultseppunct}\relax
\EndOfBibitem
\bibitem[Liland \emph{et~al.}(2010)Liland, Almøy, and Mevik]{Liland:2010}
K.~H. Liland, T.~Almøy and B.-H. Mevik, \emph{Applied Spectroscopy}, 2010, \textbf{64}, 1007--1016\relax
\mciteBstWouldAddEndPuncttrue
\mciteSetBstMidEndSepPunct{\mcitedefaultmidpunct}
{\mcitedefaultendpunct}{\mcitedefaultseppunct}\relax
\EndOfBibitem
\bibitem[Weyn \emph{et~al.}(2020)Weyn, Durran, and Caruana]{Weyn:2020}
J.~A. Weyn, D.~R. Durran and R.~Caruana, \emph{Journal of Advances in Modeling Earth Systems}, 2020, \textbf{12}, e2020MS002109\relax
\mciteBstWouldAddEndPuncttrue
\mciteSetBstMidEndSepPunct{\mcitedefaultmidpunct}
{\mcitedefaultendpunct}{\mcitedefaultseppunct}\relax
\EndOfBibitem
\bibitem[Ritvanen \emph{et~al.}(2023)Ritvanen, Harnist, Aldana, Mäkinen, and Pulkkinen]{Ritvanen:2023}
J.~Ritvanen, B.~Harnist, M.~Aldana, T.~Mäkinen and S.~Pulkkinen, \emph{IEEE Journal of Selected Topics in Applied Earth Observations and Remote Sensing}, 2023, \textbf{16}, 1654--1667\relax
\mciteBstWouldAddEndPuncttrue
\mciteSetBstMidEndSepPunct{\mcitedefaultmidpunct}
{\mcitedefaultendpunct}{\mcitedefaultseppunct}\relax
\EndOfBibitem
\bibitem[Abdalla \emph{et~al.}(2021)Abdalla, Ghaith, and Tamimi]{Abdalla:2021}
A.~M. Abdalla, I.~H. Ghaith and A.~A. Tamimi, 2021 International Conference on Information Technology (ICIT), 2021, pp. 622--626\relax
\mciteBstWouldAddEndPuncttrue
\mciteSetBstMidEndSepPunct{\mcitedefaultmidpunct}
{\mcitedefaultendpunct}{\mcitedefaultseppunct}\relax
\EndOfBibitem
\bibitem[Hamilton and Hauptmann(2018)]{Hamilton:2018}
S.~J. Hamilton and A.~Hauptmann, \emph{IEEE Transactions on Medical Imaging}, 2018, \textbf{37}, 2367--2377\relax
\mciteBstWouldAddEndPuncttrue
\mciteSetBstMidEndSepPunct{\mcitedefaultmidpunct}
{\mcitedefaultendpunct}{\mcitedefaultseppunct}\relax
\EndOfBibitem
\bibitem[Monti \emph{et~al.}(2020)Monti, Codari, van Assen, De~Cecco, and Vliegenthart]{Monti:2020}
C.~B. Monti, M.~Codari, M.~van Assen, C.~N. De~Cecco and R.~Vliegenthart, \emph{Journal of Thoracic Imaging}, 2020, \textbf{35}, S58--S65\relax
\mciteBstWouldAddEndPuncttrue
\mciteSetBstMidEndSepPunct{\mcitedefaultmidpunct}
{\mcitedefaultendpunct}{\mcitedefaultseppunct}\relax
\EndOfBibitem
\bibitem[Suganyadevi \emph{et~al.}(2022)Suganyadevi, Seethalakshmi, and Balasamy]{Suganyadevi:2022}
S.~Suganyadevi, V.~Seethalakshmi and K.~Balasamy, \emph{International Journal of Multimedia Information Retrieval}, 2022, \textbf{11}, 19--38\relax
\mciteBstWouldAddEndPuncttrue
\mciteSetBstMidEndSepPunct{\mcitedefaultmidpunct}
{\mcitedefaultendpunct}{\mcitedefaultseppunct}\relax
\EndOfBibitem
\bibitem[Sharma and Singh(2017)]{Sharma:2017}
P.~Sharma and A.~Singh, 2017 8th International Conference on Computing, Communication and Networking Technologies (ICCCNT), 2017, pp. 1--5\relax
\mciteBstWouldAddEndPuncttrue
\mciteSetBstMidEndSepPunct{\mcitedefaultmidpunct}
{\mcitedefaultendpunct}{\mcitedefaultseppunct}\relax
\EndOfBibitem
\bibitem[Shinde and Shah(2018)]{Shinde:2018}
P.~P. Shinde and S.~Shah, 2018 Fourth International Conference on Computing Communication Control and Automation (ICCUBEA), 2018, pp. 1--6\relax
\mciteBstWouldAddEndPuncttrue
\mciteSetBstMidEndSepPunct{\mcitedefaultmidpunct}
{\mcitedefaultendpunct}{\mcitedefaultseppunct}\relax
\EndOfBibitem
\bibitem[Samek \emph{et~al.}(2021)Samek, Montavon, Lapuschkin, Anders, and Müller]{Samek:2021}
W.~Samek, G.~Montavon, S.~Lapuschkin, C.~J. Anders and K.-R. Müller, \emph{Proceedings of the IEEE}, 2021, \textbf{109}, 247--278\relax
\mciteBstWouldAddEndPuncttrue
\mciteSetBstMidEndSepPunct{\mcitedefaultmidpunct}
{\mcitedefaultendpunct}{\mcitedefaultseppunct}\relax
\EndOfBibitem
\bibitem[Alzubaidi \emph{et~al.}(2021)Alzubaidi, Zhang, Humaidi, Al-Dujaili, Duan, Al-Shamma, Santamar{\'i}a, Fadhel, Al-Amidie, and Farhan]{Alzubaidi:2021}
L.~Alzubaidi, J.~Zhang, A.~J. Humaidi, A.~Al-Dujaili, Y.~Duan, O.~Al-Shamma, J.~Santamar{\'i}a, M.~A. Fadhel, M.~Al-Amidie and L.~Farhan, \emph{Journal of Big Data}, 2021, \textbf{8}, 53\relax
\mciteBstWouldAddEndPuncttrue
\mciteSetBstMidEndSepPunct{\mcitedefaultmidpunct}
{\mcitedefaultendpunct}{\mcitedefaultseppunct}\relax
\EndOfBibitem
\bibitem[Li \emph{et~al.}(2022)Li, Liu, Yang, Peng, and Zhou]{Li:2022}
Z.~Li, F.~Liu, W.~Yang, S.~Peng and J.~Zhou, \emph{IEEE Transactions on Neural Networks and Learning Systems}, 2022, \textbf{33}, 6999--7019\relax
\mciteBstWouldAddEndPuncttrue
\mciteSetBstMidEndSepPunct{\mcitedefaultmidpunct}
{\mcitedefaultendpunct}{\mcitedefaultseppunct}\relax
\EndOfBibitem
\bibitem[Liu \emph{et~al.}(2017)Liu, Osadchy, Ashton, Foster, Solomon, and Gibson]{Liu:2017}
J.~Liu, M.~Osadchy, L.~Ashton, M.~Foster, C.~J. Solomon and S.~J. Gibson, \emph{Analyst}, 2017, \textbf{142}, 4067--4074\relax
\mciteBstWouldAddEndPuncttrue
\mciteSetBstMidEndSepPunct{\mcitedefaultmidpunct}
{\mcitedefaultendpunct}{\mcitedefaultseppunct}\relax
\EndOfBibitem
\bibitem[Wahl \emph{et~al.}(2020)Wahl, Sjödahl, and Ramser]{Wahl:2020}
J.~Wahl, M.~Sjödahl and K.~Ramser, \emph{Applied Spectroscopy}, 2020, \textbf{74}, 427--438\relax
\mciteBstWouldAddEndPuncttrue
\mciteSetBstMidEndSepPunct{\mcitedefaultmidpunct}
{\mcitedefaultendpunct}{\mcitedefaultseppunct}\relax
\EndOfBibitem
\bibitem[Gebrekidan \emph{et~al.}(2021)Gebrekidan, Knipfer, and Braeuer]{Gebrekidan:2021}
M.~T. Gebrekidan, C.~Knipfer and A.~S. Braeuer, \emph{Journal of Raman Spectroscopy}, 2021, \textbf{52}, 723--736\relax
\mciteBstWouldAddEndPuncttrue
\mciteSetBstMidEndSepPunct{\mcitedefaultmidpunct}
{\mcitedefaultendpunct}{\mcitedefaultseppunct}\relax
\EndOfBibitem
\bibitem[Kazemzadeh \emph{et~al.}(2022)Kazemzadeh, Hisey, Zargar-Shoshtari, Xu, and Broderick]{Kazemzadeh:2022}
M.~Kazemzadeh, C.~L. Hisey, K.~Zargar-Shoshtari, W.~Xu and N.~G. Broderick, \emph{Optics Communications}, 2022, \textbf{510}, 127977\relax
\mciteBstWouldAddEndPuncttrue
\mciteSetBstMidEndSepPunct{\mcitedefaultmidpunct}
{\mcitedefaultendpunct}{\mcitedefaultseppunct}\relax
\EndOfBibitem
\bibitem[Luo \emph{et~al.}(2022)Luo, Popp, and Bocklitz]{Luo:2022}
R.~Luo, J.~Popp and T.~Bocklitz, \emph{Analytica}, 2022, \textbf{3}, 287--301\relax
\mciteBstWouldAddEndPuncttrue
\mciteSetBstMidEndSepPunct{\mcitedefaultmidpunct}
{\mcitedefaultendpunct}{\mcitedefaultseppunct}\relax
\EndOfBibitem
\bibitem[Valensise \emph{et~al.}(2020)Valensise, Giuseppi, Vernuccio, De~la Cadena, Cerullo, and Polli]{Valensise:2020}
C.~M. Valensise, A.~Giuseppi, F.~Vernuccio, A.~De~la Cadena, G.~Cerullo and D.~Polli, \emph{APL Photonics}, 2020, \textbf{5}, 061305\relax
\mciteBstWouldAddEndPuncttrue
\mciteSetBstMidEndSepPunct{\mcitedefaultmidpunct}
{\mcitedefaultendpunct}{\mcitedefaultseppunct}\relax
\EndOfBibitem
\bibitem[Houhou \emph{et~al.}(2020)Houhou, Barman, Schmitt, Meyer, Popp, and Bocklitz]{Houhou:2020}
R.~Houhou, P.~Barman, M.~Schmitt, T.~Meyer, J.~Popp and T.~Bocklitz, \emph{Opt. Express}, 2020, \textbf{28}, 21002--21024\relax
\mciteBstWouldAddEndPuncttrue
\mciteSetBstMidEndSepPunct{\mcitedefaultmidpunct}
{\mcitedefaultendpunct}{\mcitedefaultseppunct}\relax
\EndOfBibitem
\bibitem[Wang \emph{et~al.}(2022)Wang, O'~Dwyer, Muddiman, Ward, Camp~Jr., and Hennelly]{Wang:2022}
Z.~Wang, K.~O'~Dwyer, R.~Muddiman, T.~Ward, C.~H. Camp~Jr. and B.~M. Hennelly, \emph{Journal of Raman Spectroscopy}, 2022, \textbf{53}, 1081--1093\relax
\mciteBstWouldAddEndPuncttrue
\mciteSetBstMidEndSepPunct{\mcitedefaultmidpunct}
{\mcitedefaultendpunct}{\mcitedefaultseppunct}\relax
\EndOfBibitem
\bibitem[Junjuri \emph{et~al.}(2022)Junjuri, Saghi, Lensu, and Vartiainen]{Junjuri:2022}
R.~Junjuri, A.~Saghi, L.~Lensu and E.~M. Vartiainen, \emph{Opt. Continuum}, 2022, \textbf{1}, 1324--1339\relax
\mciteBstWouldAddEndPuncttrue
\mciteSetBstMidEndSepPunct{\mcitedefaultmidpunct}
{\mcitedefaultendpunct}{\mcitedefaultseppunct}\relax
\EndOfBibitem
\bibitem[Saghi \emph{et~al.}(2022)Saghi, Junjuri, Lensu, and Vartiainen]{Saghi:2022}
A.~Saghi, R.~Junjuri, L.~Lensu and E.~M. Vartiainen, \emph{Opt. Continuum}, 2022, \textbf{1}, 2360--2373\relax
\mciteBstWouldAddEndPuncttrue
\mciteSetBstMidEndSepPunct{\mcitedefaultmidpunct}
{\mcitedefaultendpunct}{\mcitedefaultseppunct}\relax
\EndOfBibitem
\bibitem[Junjuri \emph{et~al.}(2022)Junjuri, Saghi, Lensu, and Vartiainen]{Junjuri:2022b}
R.~Junjuri, A.~Saghi, L.~Lensu and E.~M. Vartiainen, \emph{RSC Adv.}, 2022, \textbf{12}, 28755--28766\relax
\mciteBstWouldAddEndPuncttrue
\mciteSetBstMidEndSepPunct{\mcitedefaultmidpunct}
{\mcitedefaultendpunct}{\mcitedefaultseppunct}\relax
\EndOfBibitem
\bibitem[Junjuri \emph{et~al.}(2023)Junjuri, Saghi, Lensu, and Vartiainen]{Junjuri:2023}
R.~Junjuri, A.~Saghi, L.~Lensu and E.~M. Vartiainen, \emph{Phys. Chem. Chem. Phys.}, 2023, \textbf{25}, 16340--16353\relax
\mciteBstWouldAddEndPuncttrue
\mciteSetBstMidEndSepPunct{\mcitedefaultmidpunct}
{\mcitedefaultendpunct}{\mcitedefaultseppunct}\relax
\EndOfBibitem
\bibitem[Moores \emph{et~al.}(2016)Moores, Gracie, Carson, Faulds, Graham, and Girolami]{Moores:16}
M.~T. Moores, K.~Gracie, J.~Carson, K.~Faulds, D.~Graham and M.~Girolami, 2016, arXiv:1604.07299\relax
\mciteBstWouldAddEndPuncttrue
\mciteSetBstMidEndSepPunct{\mcitedefaultmidpunct}
{\mcitedefaultendpunct}{\mcitedefaultseppunct}\relax
\EndOfBibitem
\bibitem[Härkönen \emph{et~al.}(2020)Härkönen, Roininen, Moores, and Vartiainen]{Harkonen:2020}
T.~Härkönen, L.~Roininen, M.~T. Moores and E.~M. Vartiainen, \emph{The Journal of Physical Chemistry B}, 2020, \textbf{124}, 7005--7012\relax
\mciteBstWouldAddEndPuncttrue
\mciteSetBstMidEndSepPunct{\mcitedefaultmidpunct}
{\mcitedefaultendpunct}{\mcitedefaultseppunct}\relax
\EndOfBibitem
\bibitem[Wang \emph{et~al.}(2015)Wang, Guo, Heller, and Dunson]{Wang:2015}
X.~Wang, F.~Guo, K.~A. Heller and D.~B. Dunson, Advances in Neural Information Processing Systems, 2015\relax
\mciteBstWouldAddEndPuncttrue
\mciteSetBstMidEndSepPunct{\mcitedefaultmidpunct}
{\mcitedefaultendpunct}{\mcitedefaultseppunct}\relax
\EndOfBibitem
\bibitem[Magris and Iosifidis(2023)]{Magris:2023}
M.~Magris and A.~Iosifidis, \emph{Artificial Intelligence Review}, 2023,  1--51\relax
\mciteBstWouldAddEndPuncttrue
\mciteSetBstMidEndSepPunct{\mcitedefaultmidpunct}
{\mcitedefaultendpunct}{\mcitedefaultseppunct}\relax
\EndOfBibitem
\bibitem[Abdar \emph{et~al.}(2021)Abdar, Pourpanah, Hussain, Rezazadegan, Liu, Ghavamzadeh, Fieguth, Cao, Khosravi, Acharya, Makarenkov, and Nahavandi]{Abdar:2021}
M.~Abdar, F.~Pourpanah, S.~Hussain, D.~Rezazadegan, L.~Liu, M.~Ghavamzadeh, P.~Fieguth, X.~Cao, A.~Khosravi, U.~R. Acharya, V.~Makarenkov and S.~Nahavandi, \emph{Information Fusion}, 2021, \textbf{76}, 243--297\relax
\mciteBstWouldAddEndPuncttrue
\mciteSetBstMidEndSepPunct{\mcitedefaultmidpunct}
{\mcitedefaultendpunct}{\mcitedefaultseppunct}\relax
\EndOfBibitem
\bibitem[Cui \emph{et~al.}(2016)Cui, Law, and Marzouk]{Cui:2016}
T.~Cui, K.~J. Law and Y.~M. Marzouk, \emph{Journal of Computational Physics}, 2016, \textbf{304}, 109--137\relax
\mciteBstWouldAddEndPuncttrue
\mciteSetBstMidEndSepPunct{\mcitedefaultmidpunct}
{\mcitedefaultendpunct}{\mcitedefaultseppunct}\relax
\EndOfBibitem
\bibitem[Morzfeld \emph{et~al.}(2019)Morzfeld, Tong, and Marzouk]{Morzfeld:2019}
M.~Morzfeld, X.~Tong and Y.~Marzouk, \emph{Journal of Computational Physics}, 2019, \textbf{380}, 1--28\relax
\mciteBstWouldAddEndPuncttrue
\mciteSetBstMidEndSepPunct{\mcitedefaultmidpunct}
{\mcitedefaultendpunct}{\mcitedefaultseppunct}\relax
\EndOfBibitem
\bibitem[Yang \emph{et~al.}(2023)Yang, Robeyns, Wang, and Aitchison]{Yang:2023}
A.~X. Yang, M.~Robeyns, X.~Wang and L.~Aitchison, 2023, arXiv:2308.13111\relax
\mciteBstWouldAddEndPuncttrue
\mciteSetBstMidEndSepPunct{\mcitedefaultmidpunct}
{\mcitedefaultendpunct}{\mcitedefaultseppunct}\relax
\EndOfBibitem
\bibitem[Sharma \emph{et~al.}(2023)Sharma, Farquhar, Nalisnick, and Rainforth]{Sharma:2023}
M.~Sharma, S.~Farquhar, E.~Nalisnick and T.~Rainforth, 2023, arXiv:2211.06291\relax
\mciteBstWouldAddEndPuncttrue
\mciteSetBstMidEndSepPunct{\mcitedefaultmidpunct}
{\mcitedefaultendpunct}{\mcitedefaultseppunct}\relax
\EndOfBibitem
\bibitem[Antonio \emph{et~al.}(2023)Antonio, O'Toole, Carney, Kulkarni, and Palazoglu]{Antonio:2023}
D.~Antonio, H.~O'Toole, R.~Carney, A.~Kulkarni and A.~Palazoglu, 2023, arXiv:2306.16621\relax
\mciteBstWouldAddEndPuncttrue
\mciteSetBstMidEndSepPunct{\mcitedefaultmidpunct}
{\mcitedefaultendpunct}{\mcitedefaultseppunct}\relax
\EndOfBibitem
\bibitem[Mozaffari and Tay(2021)]{Mozaffari:2021}
M.~H. Mozaffari and L.-L. Tay, 2021 5th SLAAI International Conference on Artificial Intelligence (SLAAI-ICAI), 2021, pp. 1--6\relax
\mciteBstWouldAddEndPuncttrue
\mciteSetBstMidEndSepPunct{\mcitedefaultmidpunct}
{\mcitedefaultendpunct}{\mcitedefaultseppunct}\relax
\EndOfBibitem
\bibitem[Rasmussen and Williams(2005)]{Rasmussen:2005}
C.~E. Rasmussen and C.~K.~I. Williams, \emph{Gaussian Processes for Machine Learning}, MIT Press, 2005\relax
\mciteBstWouldAddEndPuncttrue
\mciteSetBstMidEndSepPunct{\mcitedefaultmidpunct}
{\mcitedefaultendpunct}{\mcitedefaultseppunct}\relax
\EndOfBibitem
\bibitem[Härkönen \emph{et~al.}(2023)Härkönen, Hannula, Moores, Vartiainen, and Roininen]{HarkonenFoDS:2023}
T.~Härkönen, E.~Hannula, M.~T. Moores, E.~M. Vartiainen and L.~Roininen, \emph{Foundations of Data Science}, 2023,  DOI: 10.3934/fods.2023008\relax
\mciteBstWouldAddEndPuncttrue
\mciteSetBstMidEndSepPunct{\mcitedefaultmidpunct}
{\mcitedefaultendpunct}{\mcitedefaultseppunct}\relax
\EndOfBibitem
\bibitem[M{\o}ller \emph{et~al.}(1998)M{\o}ller, Syversveen, and Waagepetersen]{Moller:1998}
J.~M{\o}ller, A.~R. Syversveen and R.~P. Waagepetersen, \emph{Scandinavian Journal of Statistics}, 1998, \textbf{25}, 451--482\relax
\mciteBstWouldAddEndPuncttrue
\mciteSetBstMidEndSepPunct{\mcitedefaultmidpunct}
{\mcitedefaultendpunct}{\mcitedefaultseppunct}\relax
\EndOfBibitem
\bibitem[Hilbe(2011)]{Hilbe:2011}
J.~M. Hilbe, \emph{Negative Binomial Regression}, Cambridge University Press, $2^\text{nd}$ edn, 2011\relax
\mciteBstWouldAddEndPuncttrue
\mciteSetBstMidEndSepPunct{\mcitedefaultmidpunct}
{\mcitedefaultendpunct}{\mcitedefaultseppunct}\relax
\EndOfBibitem
\bibitem[Teng \emph{et~al.}(2017)Teng, Nathoo, and Johnson]{Teng:2017}
M.~Teng, F.~Nathoo and T.~D. Johnson, \emph{Journal of Statistical Computation and Simulation}, 2017, \textbf{87}, 2227--2252\relax
\mciteBstWouldAddEndPuncttrue
\mciteSetBstMidEndSepPunct{\mcitedefaultmidpunct}
{\mcitedefaultendpunct}{\mcitedefaultseppunct}\relax
\EndOfBibitem
\bibitem[Haario \emph{et~al.}(2006)Haario, Laine, Mira, and Saksman]{Haario:2006}
H.~Haario, M.~Laine, A.~Mira and E.~Saksman, \emph{Statistics and Computing}, 2006, \textbf{16}, 339--354\relax
\mciteBstWouldAddEndPuncttrue
\mciteSetBstMidEndSepPunct{\mcitedefaultmidpunct}
{\mcitedefaultendpunct}{\mcitedefaultseppunct}\relax
\EndOfBibitem
\bibitem[Srivastava \emph{et~al.}(2014)Srivastava, Hinton, Krizhevsky, Sutskever, and Salakhutdinov]{Srivastava:2014}
N.~Srivastava, G.~Hinton, A.~Krizhevsky, I.~Sutskever and R.~Salakhutdinov, \emph{Journal of Machine Learning Research}, 2014, \textbf{15}, 1929--1958\relax
\mciteBstWouldAddEndPuncttrue
\mciteSetBstMidEndSepPunct{\mcitedefaultmidpunct}
{\mcitedefaultendpunct}{\mcitedefaultseppunct}\relax
\EndOfBibitem
\bibitem[Wilson \emph{et~al.}(2016)Wilson, Hu, Salakhutdinov, and Xing]{Wilson:2016}
A.~G. Wilson, Z.~Hu, R.~Salakhutdinov and E.~P. Xing, Proceedings of the 19th International Conference on Artificial Intelligence and Statistics, Cadiz, Spain, 2016, pp. 370--378\relax
\mciteBstWouldAddEndPuncttrue
\mciteSetBstMidEndSepPunct{\mcitedefaultmidpunct}
{\mcitedefaultendpunct}{\mcitedefaultseppunct}\relax
\EndOfBibitem
\bibitem[Calandra \emph{et~al.}(2016)Calandra, Peters, Rasmussen, and Deisenroth]{Calandra:2016}
R.~Calandra, J.~Peters, C.~E. Rasmussen and M.~P. Deisenroth, 2016 International joint conference on neural networks (IJCNN), 2016, pp. 3338--3345\relax
\mciteBstWouldAddEndPuncttrue
\mciteSetBstMidEndSepPunct{\mcitedefaultmidpunct}
{\mcitedefaultendpunct}{\mcitedefaultseppunct}\relax
\EndOfBibitem
\bibitem[Abadi \emph{et~al.}(2015)Abadi, Agarwal, Barham, Brevdo, Chen, Citro, Corrado, Davis, Dean, Devin, Ghemawat, Goodfellow, Harp, Irving, Isard, Jia, Jozefowicz, Kaiser, Kudlur, Levenberg, Man\'{e}, Monga, Moore, Murray, Olah, Schuster, Shlens, Steiner, Sutskever, Talwar, Tucker, Vanhoucke, Vasudevan, Vi\'{e}gas, Vinyals, Warden, Wattenberg, Wicke, Yu, and Zheng]{Tensorflow:2015}
M.~Abadi, A.~Agarwal, P.~Barham, E.~Brevdo, Z.~Chen, C.~Citro, G.~S. Corrado, A.~Davis, J.~Dean, M.~Devin, S.~Ghemawat, I.~Goodfellow, A.~Harp, G.~Irving, M.~Isard, Y.~Jia, R.~Jozefowicz, L.~Kaiser, M.~Kudlur, J.~Levenberg, D.~Man\'{e}, R.~Monga, S.~Moore, D.~Murray, C.~Olah, M.~Schuster, J.~Shlens, B.~Steiner, I.~Sutskever, K.~Talwar, P.~Tucker, V.~Vanhoucke, V.~Vasudevan, F.~Vi\'{e}gas, O.~Vinyals, P.~Warden, M.~Wattenberg, M.~Wicke, Y.~Yu and X.~Zheng, 2015, arXiv:1603.04467\relax
\mciteBstWouldAddEndPuncttrue
\mciteSetBstMidEndSepPunct{\mcitedefaultmidpunct}
{\mcitedefaultendpunct}{\mcitedefaultseppunct}\relax
\EndOfBibitem
\bibitem[Dillon \emph{et~al.}(2017)Dillon, Langmore, Tran, Brevdo, Vasudevan, Moore, Patton, Alemi, Hoffman, and Saurous]{Dillon:2017}
J.~V. Dillon, I.~Langmore, D.~Tran, E.~Brevdo, S.~Vasudevan, D.~Moore, B.~Patton, A.~Alemi, M.~Hoffman and R.~A. Saurous, 2017, arXiv:1711.10604\relax
\mciteBstWouldAddEndPuncttrue
\mciteSetBstMidEndSepPunct{\mcitedefaultmidpunct}
{\mcitedefaultendpunct}{\mcitedefaultseppunct}\relax
\EndOfBibitem
\bibitem[Chollet \emph{et~al.}(2015)Chollet\emph{et~al.}]{Chollet:2015}
F.~Chollet \emph{et~al.}, \emph{Keras}, \url{https://keras.io}, 2015\relax
\mciteBstWouldAddEndPuncttrue
\mciteSetBstMidEndSepPunct{\mcitedefaultmidpunct}
{\mcitedefaultendpunct}{\mcitedefaultseppunct}\relax
\EndOfBibitem
\bibitem[ram()]{ramanDatabase}
\emph{{The standard Pigments Checker v.5}}, \url{https://chsopensource.org/pigments-checker/}, Accessed: 2023-02-14\relax
\mciteBstWouldAddEndPuncttrue
\mciteSetBstMidEndSepPunct{\mcitedefaultmidpunct}
{\mcitedefaultendpunct}{\mcitedefaultseppunct}\relax
\EndOfBibitem
\bibitem[Jiang \emph{et~al.}(2017)Jiang, Wu, Zheng, and Wong]{Jiang:2017}
B.~Jiang, T.-Y. Wu, C.~Zheng and W.~H. Wong, \emph{Statistica Sinica}, 2017, \textbf{27}, 1595--1618\relax
\mciteBstWouldAddEndPuncttrue
\mciteSetBstMidEndSepPunct{\mcitedefaultmidpunct}
{\mcitedefaultendpunct}{\mcitedefaultseppunct}\relax
\EndOfBibitem
\bibitem[Grazian and Fan(2020)]{Grazian:2020}
C.~Grazian and Y.~Fan, \emph{WIREs Computational Statistics}, 2020, \textbf{12}, e1486\relax
\mciteBstWouldAddEndPuncttrue
\mciteSetBstMidEndSepPunct{\mcitedefaultmidpunct}
{\mcitedefaultendpunct}{\mcitedefaultseppunct}\relax
\EndOfBibitem
\bibitem[Tresp(2001)]{tresp2001mixtures}
V.~Tresp, \emph{Advances in Neural Information Processing Systems}, 2001,  654--660\relax
\mciteBstWouldAddEndPuncttrue
\mciteSetBstMidEndSepPunct{\mcitedefaultmidpunct}
{\mcitedefaultendpunct}{\mcitedefaultseppunct}\relax
\EndOfBibitem
\bibitem[Rasmussen and Ghahramani(2002)]{Rasmussen:2002}
C.~E. Rasmussen and Z.~Ghahramani, \emph{Advances in Neural Information Processing Systems 14}, MIT Press, 2002, pp. 881--888\relax
\mciteBstWouldAddEndPuncttrue
\mciteSetBstMidEndSepPunct{\mcitedefaultmidpunct}
{\mcitedefaultendpunct}{\mcitedefaultseppunct}\relax
\EndOfBibitem
\bibitem[Härkönen \emph{et~al.}(2022)Härkönen, Wade, Law, and Roininen]{Harkonen:2022:MoE}
T.~Härkönen, S.~Wade, K.~Law and L.~Roininen, \emph{Mixtures of {G}aussian Process Experts with {SMC}$^2$}, 2022, arXiv:2208.12830\relax
\mciteBstWouldAddEndPuncttrue
\mciteSetBstMidEndSepPunct{\mcitedefaultmidpunct}
{\mcitedefaultendpunct}{\mcitedefaultseppunct}\relax
\EndOfBibitem
\bibitem[{Van Wittenberghe} \emph{et~al.}(2014){Van Wittenberghe}, Verrelst, Rivera, Alonso, Moreno, and Samson]{Wittenberghe:2014}
S.~{Van Wittenberghe}, J.~Verrelst, J.~P. Rivera, L.~Alonso, J.~Moreno and R.~Samson, \emph{Journal of Photochemistry and Photobiology B: Biology}, 2014, \textbf{134}, 37--48\relax
\mciteBstWouldAddEndPuncttrue
\mciteSetBstMidEndSepPunct{\mcitedefaultmidpunct}
{\mcitedefaultendpunct}{\mcitedefaultseppunct}\relax
\EndOfBibitem
\bibitem[Lázaro-Gredilla \emph{et~al.}(2014)Lázaro-Gredilla, Titsias, Verrelst, and Camps-Valls]{Lazaro-Gredilla:2014}
M.~Lázaro-Gredilla, M.~K. Titsias, J.~Verrelst and G.~Camps-Valls, \emph{IEEE Geoscience and Remote Sensing Letters}, 2014, \textbf{11}, 838--842\relax
\mciteBstWouldAddEndPuncttrue
\mciteSetBstMidEndSepPunct{\mcitedefaultmidpunct}
{\mcitedefaultendpunct}{\mcitedefaultseppunct}\relax
\EndOfBibitem
\bibitem[Ghosh \emph{et~al.}(2021)Ghosh, Li, Zeng, Zhang, and Zhou]{Ghosh:2021}
M.~Ghosh, Y.~Li, L.~Zeng, Z.~Zhang and Q.~Zhou, \emph{IISE Transactions}, 2021, \textbf{53}, 787--798\relax
\mciteBstWouldAddEndPuncttrue
\mciteSetBstMidEndSepPunct{\mcitedefaultmidpunct}
{\mcitedefaultendpunct}{\mcitedefaultseppunct}\relax
\EndOfBibitem
\bibitem[Lawrence \emph{et~al.}(2006)Lawrence, Sanguinetti, and Rattray]{Lawrence:2006}
N.~Lawrence, G.~Sanguinetti and M.~Rattray, Advances in Neural Information Processing Systems, 2006\relax
\mciteBstWouldAddEndPuncttrue
\mciteSetBstMidEndSepPunct{\mcitedefaultmidpunct}
{\mcitedefaultendpunct}{\mcitedefaultseppunct}\relax
\EndOfBibitem
\bibitem[Bu \emph{et~al.}(2020)Bu, Kumar, Xie, Pan, Zhao, and Wu]{Bu:2022}
Y.~Bu, Y.~B. Kumar, J.~Xie, J.~Pan, G.~Zhao and Y.~Wu, \emph{The Astrophysical Journal Supplement Series}, 2020, \textbf{249}, 7\relax
\mciteBstWouldAddEndPuncttrue
\mciteSetBstMidEndSepPunct{\mcitedefaultmidpunct}
{\mcitedefaultendpunct}{\mcitedefaultseppunct}\relax
\EndOfBibitem
\bibitem[Härkönen \emph{et~al.}(2023)Härkönen, Sundström, Tamminen, Hakkarainen, Vakkilainen, and Haario]{Harkonen:2023:Plume}
T.~Härkönen, A.-M. Sundström, J.~Tamminen, J.~Hakkarainen, E.~Vakkilainen and H.~Haario, \emph{International Journal for Uncertainty Quantification}, 2023, \textbf{13}, 41--59\relax
\mciteBstWouldAddEndPuncttrue
\mciteSetBstMidEndSepPunct{\mcitedefaultmidpunct}
{\mcitedefaultendpunct}{\mcitedefaultseppunct}\relax
\EndOfBibitem
\bibitem[Gal and Ghahramani(2016)]{Gal:2016}
Y.~Gal and Z.~Ghahramani, Proceedings of The 33rd International Conference on Machine Learning, New York, New York, USA, 2016, pp. 1050--1059\relax
\mciteBstWouldAddEndPuncttrue
\mciteSetBstMidEndSepPunct{\mcitedefaultmidpunct}
{\mcitedefaultendpunct}{\mcitedefaultseppunct}\relax
\EndOfBibitem
\bibitem[Talts \emph{et~al.}(2020)Talts, Betancourt, Simpson, Vehtari, and Gelman]{Talts:2020}
S.~Talts, M.~Betancourt, D.~Simpson, A.~Vehtari and A.~Gelman, 2020, arXiv:1804.06788\relax
\mciteBstWouldAddEndPuncttrue
\mciteSetBstMidEndSepPunct{\mcitedefaultmidpunct}
{\mcitedefaultendpunct}{\mcitedefaultseppunct}\relax
\EndOfBibitem
\end{mcitethebibliography}
\bibliographystyle{rsc} %the RSC's .bst file

\end{document}